\documentclass[manuscript]{aastex}

\slugcomment{To appear in the November 2008 {\it Astronomical Journal}.}

\shorttitle{QSO {\it K} Corrections}
\shortauthors{Kennefick\&Bursick}

\received{2007 July 6}
\accepted{2008 August 14}
\begin{document}

\title{Infrared Imaging of SDSS Quasars:\\
    Implications for the Quasar {\it K} correction}


\author{Julia Kennefick\altaffilmark{1,2,3,4} and Shelly Bursick\altaffilmark{1,2,3}}
\email{jkennef@uark.edu}


\altaffiltext{1}{Department of Physics, University of Arkansas, 226 Physics Building,
	Fayetteville, AR 72701.} 
\altaffiltext{2}{Arkansas Center for Space and Planetary Sciences,
       University of Arkansas, 202 Old Museum Building, Fayetteville, AR 72701} 
\altaffiltext{3}{Visiting Astronomer, Kitt Peak National Observatory.
	KPNO is operated by AURA, Inc.\ under contract to the National Science
	Foundation.}
\altaffiltext{4}{NSF ADVANCE Fellow}


\begin{abstract}
We have imaged 45 quasars from the Sloan Digital Sky Survey (SDSS) with redshifts
$1.85 < z < 4.26$ in $JHK_s$ with the KPNO SQIID imager. By combining these
data with optical magnitudes from the SDSS we have computed the restframe
optical spectral indices of this sample and investigate their 
relation to quasar redshift.  We find a mean spectral index of
$\langle\alpha_o\rangle = -0.55\pm0.42$ with a large spread in values.
We also find possible evolution of the form
$\alpha_o = (0.148\pm0.068)z - (0.964\pm0.200)$ in the luminosity range
$-28.0 < M_i < -26.5$.
Such evolution suggests changes in the accretion process in quasars with
time and is shown to have an effect on computed quasar luminosity functions.

\end{abstract}



\section{Introduction}

The recent publication of large samples of quasars 
from the Sloan Digital Sky Survey \citep[SDSS;][]{sch07} and the Two Degree
Field QSO Redshift Survey \citep[2QZ;][]{cro04} allows
the study of the evolution of the quasar population spanning long lookback times
with large, homogenous catalogs of objects and the characterization of the 
quasar luminosity function (QLF).   
The \citet{ric06a} QLF formed from the SDSS Third Data Release Quasar Catalog
(DR3QLF) is defined from 15,343 quasars over 1622 deg$^2$,
spanning redshifts from $z=0$ to $5$ and reaching luminosities down to $M_i < -23$ in 
their lowest redshift bins.  They find a peak in type 1 quasar activity between
$z=2.2$ and $2.8$ and a general flattening of the slope of the bright-end QLF 
with increasing redshift. \citet{hop07} compute a bolmetric QLF, combining many
quasars surveys, including the DR3QLF and the 2QZ.  They find a peak in
the quasar luminosity density at $z=2.15$ and an evolving QLF bright-end slope 
that becomes flatter at redshifts $z>3$.

One major difficulty in characterizing the evolution in quasar space densities 
is finding an appropriate way to calculate the absolute magnitudes of a 
survey sample and compare these magnitudes with other surveys, or even to 
objects within the same survey if it spans a large redshift range, i.e., how 
to calculate the appropriate {\it K} correction.  Traditionally, absolute magnitudes
were corrected back to the restframe $B$-band \citep{sch83,boy87}, as most early 
surveys for quasars included the measurement of flux around $4400\AA$, by assuming
a power-law form for quasar spectra redward of Lyman-$\alpha$ of the form
$f_\nu \propto \nu^{\alpha}$ and assuming an average spectral index, $\alpha$, for
the survey sample.  As surveys
move to higher redshifts, beyond $z=3$, the observed $B$-band ceases to sample the quasar
continuum effectively, as the Lyman-$\alpha$ line moves into and redward of that
filter.  In the recent past, research groups have adopted several methods for 
computing absolute magnitudes and have referenced their magnitudes to different
restframe wavelengths.  
For instance, \citet{sch95} and \citet{ken95} adopt an average spectral index of 
$\alpha=-0.5$ in their computations of $M_B$, while \citet{war94} avoid the adoption 
of a spectral
index by computing the QLF as a function of $M_{C(1216)}$, the flux in the continuum
under the Lyman-$\alpha$ line, from their direct measurements of the flux at that
point in their survey spectra.  More recently, \citet{cro04} present their QLF from
the 2QZ at $0.4 < z < 2.1$ in terms of $M_{b_J}$, and \citet{ric06a} assume an average
spectral index, but correct $i$-band magnitudes to $z=2$.   The correction to $z=2$
is prompted by the desire to minimize the effects of extrapolating the assumed quasar
powerlaw over large wavelengths, as suggested by \citet{wis00}.

Such differences in the methods for computing absolute magnitudes has led to some 
discrepancies in computed quasar space densities in past surveys \citep{ken97}.
The initial goals of this near-infrared (NIR) imaging program were to measure the 
restframe $B$-band flux
for a set of quasars in several redshift ranges, to compare the differences in the
absolute magnitudes, $M_B$, as computed from extrapolations of flux values at
lower wavelengths to flux values measured directly, and to see how this changed
with redshift.  This is equivalent to measuring the restframe optical spectral
index of the quasars from optical and NIR photometry.  In this paper we report
the initial results of a program to measure the restframe optical spectral index, $\alpha_o$,
of a subset of SDSS quasars at $1.85 < z < 4.26$ using their reported optical 
magnitudes and NIR photometry obtained at the KPNO 2.1m telescope using 
the Simultaneous Quad Infrared Imaging Device (SQIID).
In \S 2 we present the program design.  The data used in the project, both archived
and new observations, are described in \S 3. In \S 4 we give the program results, 
including computations of spectral indices and the implications for the 
resulting absolute magnitudes.  We discuss our results further in \S 5.  

\section{Program Design}
The initial goal of this observing program was to measure the restframe
$B$-band flux for a sample of quasars at a range of redshifts 
in order to essentially bypass the need to extrapolate a power-law over large 
wavelength ranges when applying a {\it K} correction in the computation of quasar
absolute magnitudes, $M_B$.  
The central wavelength of the $J$-band is 1.27 microns, which corresponds to the
restframe $B$-band (central wavelength of 4400 $\AA$)
at $z=1.88$.  Likewise, $H$-band corresponds to the restframe $B$-band 
at $z=2.80$, and $K_s$-band at $z=4.06$.  Figure 1 shows the NIR $JHK_s$
filter curves, along with the ``redshifted'' Johnson $B$-band at these
redshifts, that is, the portion of the spectrum you would like to measure to 
sample the restframe $B$-band at the given redshift.  For this project, we targeted 
45 quasars chosen from the SDSS quasar catalogs for NIR imaging centered around the 
above redshift ranges.  We combine this NIR photometry with reported optical 
magnitudes to compute restframe optical spectral indices for the sample.   
Programatically, this can be achieved by fitting the photometric data to measure 
the slope of the spectral energy distribution, $\alpha$, as dicussed further below.    

\citet{ric06a} has reported the DR3QLF computed from SDSS quasars as a function 
of $M_i$ corrected to a redshift of $z=2$, and the selection of the $B$-band
for computations of the QLF has become less common as quasar surveys move to higher
redshift.  Here, we compute the optical spectral index, $\alpha_o$, for a subset of 
the SDSS quasars using optical and NIR photometry.  
We then explore possible correlations of $\alpha_o$ with redshift or luminosity
and how that might affect the characterization of the QLF and its evolution.

\section{Observations and Data Reductions}

\subsection{SQIID Near-Infrared Imaging}

Forty-five quasars from the SDSS Third Data Release Quasar Catalog 
\citep[DR3Q;][]{sch05} were imaged with SQIID 
on the KPNO 2.1m telescope during 2005 March 22-23.  The SQIID Infrared Camera
uses individual $512 \times 512$ pixel quadrants of {\textsc ALADDIN} InSb arrays, 
with a pixel scale of 0.69\arcsec pixel$^{-1}$ at the KPNO 2.1m.  The effective 
field of view is 304\arcsec $\times$ 317\arcsec.  We configured SQIID to take 
images in the $JHK_s$ bands, which it acquires simultaneously by splitting the beam 
with a series of dichroics before sending the beam to separate NIR cameras
optimized to perform over a limited wavelength range.  

The observing strategy involved imaging each target quasar for a total of 600s, 
with five pointings of 120s, each the sum of fifteen 8s coadds, offset by roughly
45\arcsec\ between pointings to improve background subtraction.
Conditions were nonphotometric, with light cirrus
over the course of the run.  Seeing ranged from 1.3\arcsec\ to 1.9\arcsec\ over the
two nights of the observing run.   

The images were processed using the {\textsc IRAF}\footnote{\textsc{IRAF} is 
distributed by the National Optical Astronomy Observatory, which is operated by
the AURA, Inc under cooperative agreement with the NSF.} 
{\textsc UPSQIID}\footnote{http://www.noao.edu/kpno/sqiid/} package.
Dark frames were acquired for each combination of coadd and exposure time 
utilized during the run.   
A global flatfield for each night and each filter 
was constructed from the object frames using the 
{\textsc USQFLAT} routine, configured to subtract the dark current
and then to perform a median combine of the frames. 
Image processing was performed using the 
{\textsc MOVPROC} routine to generate and subtract a moving sky image created from
the 6 frames obtained closest to the object frame in time and to correct for pixel to
pixel sensitivity differences by dividing by the appropriate flatfield.   
The final image of each quasar in each band was constructed by registering 
the five pointings using the {\textsc USQMOS} routines and combining the frames
using the {\textsc NIRCOMBINE} routine.  

Object detection and measurement was carried out using the Source 
Extractor\footnote{http://terapix.iap.fr/} software. The observations
were made under fair but nonphotometric conditions. Calibration was
performed through the use of stellar Two Micron All Sky Survey \citep[2MASS;][]{cut03}
sources in the object
frames. The 2MASS images and catalogs were accessed through the 
National Virtual Observatory\footnote{http://www.us-vo.org/} (NVO)
Open SkyQuery and DataScope Query tools and manipulated using the Aladin multiview tool
\citep{bon00}.  The $JHK_s$ magnitudes reported in Table 1 and used in the
following analysis were derived from the flux inside circular apertures of
radius 3.5\arcsec.  The photometric zeropoints for the frames were calculated
using stellar 2MASS sources in the frame of the quasar with measured $JHK_S$ aperture
magnitudes.  Typically five to seven calibration stars were used with 
magnitudes ranging from 14 to 16 in $J$ and $H$, and from 14 to 15 in $K_s$.  The 2MASS 
aperture magnitudes were measured in a 4\arcsec\ radius and ``curve-of-growth'' 
corrected out to an ``infinite''
aperture\footnote{http://www.ipac.caltech.edu/2mass/releases/allsky/doc/}. 
Therefore, the magnitudes given in Table 1 are effectively aperture 
corrected by using these corrected 2MASS sources as standards.
Near-infrared magnitudes, uncorrected for Galactic extinction, for the quasars are 
given in Table 1 and Figure 2, along with their associated errors, which include 
both measured photometric errors for the quasars and the errors introduced by the 
calibration process.   
The quasars in the subsample observed with SQIID are 1 to 2 magnitudes fainter than
the SDSS quasars detected by 2MASS.  The relatively few numbers of quasars in the 
highest redshift bin ($\sim4.06$) meant that we had to choose fainter objects over
a slightly wider redshift range than in the two lower redshift bins.   

\subsection{The SDSS Third Data Release Quasar Catalog}
The SDSS uses a multicolor technique to select
quasar candidates in the optical $ugriz$ bands over the redshift range
$0.08 < z < 5.41$.  The DR3Q contains 46,420 quasars with luminosities 
$M_i > -22$.
Initially, we chose 60 quasars from the SDSS Early Data Release 
quasar catalog \citep[EDR;][]{sch02} for NIR imaging. 
Quasars were selected from the EDR catalog randomly from those 
objects with r.a. between $9^h$ and $16^h$ and as close to the target
redshifts as possible.  Since the EDR contains more objects at lower
redshifts, there is less spread in the sample redshifts at $z=1.88$,
increasing slightly at $z=2.80$ and more pronounced at $z=4.06$ (see Figures 2 and 3).   
The mean redshifts for the three bins are $z=1.88$, $2.82$, and $4.03$.
We obtained NIR imaging for 45 of these objects, chosen to sample our redshift
ranges and with RA's close to the local sidereal time during the observations. 
All of the imaged quasars are contained in the SDSS DR3Q catalog,
and the optical photometry reported in Table 1 and used in the
data analysis were taken from this later catalog.  

The SDSS DR3Q has been matched to the 2MASS All-Sky Data Release Point Source 
Catalog \citep{cut03} and the 2MASS $JHK_s$ magnitudes and errors are reported 
for those SDSS quasars with 2MASS detections (columns 25-30 of the DR3Q.)  
We have used the 6192 quasars with 2MASS detections as a bright comparison sample  
to our fainter (1$^m$ to 1.5$^m$) SQIID sample (Figures 2 and 3).   

\section{Results}
\subsection{Colors}
The magnitudes were corrected for Galactic extinction using the value of $A_u$ for each
object reported in the SDSS DR3Q catalog, column 15 \citep{sch05}.  These values were
taken from the maps of \citet{sch98} which assumes an extinction to reddening ratio $R_V = 3.1$.
Extinctions in the Sloan $griz$ bands are then 0.736, 0.534, 0.405, and 0.287 times $A_u$,
respectively.  The value of $A/E(B-V)$ for the Sloan $u$-band is 5.155;  consulting \citet{sch98}
for the UKIRT $JHK$ values gives 0.902, 0.576, and 0.367, respectively.  This gives values for
the extinction in these bands of $A_J = 0.175A_u$, $A_H = 0.112A_u$, and $A_K = 0.071A_u$.
Colors computed from these extinction corrected magnitudes for the SQIID sample and the SDSS
quasars detected by 2MASS are shown in Figure 3.  The SQIID sample spans the same
range in quasar colors as do the 2MASS detected quasars.  

Also shown in Figure 3 are expected colors for quasars computed using
synthetic quasar spectra generated with power law spectral indices and
averaged over several different realizations of the Lyman-$\alpha$ forest.
The colors were computed by passing the quasar spectra through 
SDSS and SQIID filters and calibrated with a comparison Vega spectrum
from the Space Telescope Science Institute
{\textsc CALSPEC}\footnote{http://www.stsci.edu/hst/cdbs/calspec.html} site.
The $ugriz$ Vega based magnitudes were then converted to the SDSS $AB$ system using \citet{hb06}.  
The dotted line in Figure 3 represents quasars with $\alpha=-0.5$. 
While the quasar colors do cluster around this line, the colors
span a considerable range, larger than the errors in their colors.  This is consistent
with results found by \citet{pen03}, who found a similar spread in optical/NIR
colors in a sample of 45 SDSS quasars at $3.6 < z < 5.03$.  

\subsection{Optical Spectral Indices}
In order to compute the restframe optical continuum slopes of the quasar spectral
energy distributions, we have converted our NIR, Vega
based photometry to the AB system.  To accomplish this, we first constructed a NIR spectrum for
Vega based on the absolute flux calibrations for Vega reported by \citet{meg95} (see their Table 4,)
constraining the blue end to have $f_\nu = 2446 JY$ at $\lambda = 9000\AA$, as given in their 
formula for the blackbody fit to Vega (see their page 776.)
We then compared the flux from this spectrum with a flat source with $f_\nu = 3631 Jy$, the 
zeropoint of the AB system \citep{og83}, for each 2MASS and SQIID $JHK_s$ filter.  The offsets
for the 2MASS filter set are:  $J_{AB} = J + 0.84$, $H_{AB} = H + 1.33$, and $K_{AB} = K_s + 1.79$.
For the SQIID filter set they are: $J_{AB} = J + 0.88$, $H_{AB} = H + 1.35$, and 
$K_{AB} = K_s + 1.75$.

If we assume that the optical flux for a quasar is given by
$f_\nu = \nu^{\alpha_o}$, then we can rewrite the formula for the AB magnitude \citep{og83} 
\begin{equation}
AB = -2.5\log_{10} f_\nu - 48.6
\end{equation}
in the following form:
\begin{equation}
AB = -2.5\alpha_o\log_{10} \nu - 48.6,
\end{equation}
allowing us to compute $\alpha_o$ as the slope of a fit to straight line in
AB magnitude vs. $\log\nu$.  This was accomplished by using the Numerical Recipes
routine {\textsc FIT} \citep{nr92}.
Spectral indices for each quasar observed with SQIID are given
in Table 2. 
The mean spectral index for the sample is $\langle\alpha_o\rangle = -0.55\pm 0.42$. 
However, there appears to be a change in the mean slope with 
redshift in the SQIID sample, with
quasars at lower redshifts having a steeper slope.  
In Figure 4, AB magnitudes transformed to the quasar restframe
and normalized to 20.0 in the bluest band completely redward of Lyman-$\alpha$
are given along with a line showing the mean slope for that redshift range. 
The mean spectral index for each of the redshift bins
is:  $\langle\alpha_o\rangle_{1.88} = -0.71\pm0.43$, 
$\langle\alpha_o\rangle_{2.82} = -0.49\pm0.40$,
$\langle\alpha_o\rangle_{4.03} = -0.40\pm0.39$ (see also Table 3).  
These spectral indices were calculated using
all available passbands.  For consistency, if we only use those passbands available
in all three redshift bins (from restframe $\sim1500\AA$ out to $\sim4500\AA)$, the computed
spectral indices change very little, with means of:  $\langle\alpha_o\rangle_{1.88} =
-0.68$, $\langle\alpha_o\rangle_{2.82} = -0.50$, $\langle\alpha_o\rangle_{4.03} = -0.40$.

\subsection{{\it K} corrections and Absolute Magnitudes}
In order to compute a luminosity function for quasars, the absolute magnitude
of each discovered object must be calculated.  This can be a bolometric
luminosity, the luminosity at a point in the continuum, or the
flux in a passband.  However, since the spectra of quasars are redshifted
and even an individual survey can span a large range in redshift, the
calculation must include a conversion from an observed band to an emitted band.
This conversion from observed flux to restframe flux is referred to as 
the {\it K} correction \citep{hum56}, defined as the 
``technical effect that occurs when a continuous energy
distribution $F(\lambda)$ is redshifted through {\it fixed} spectral-response bands
of a detector'' \citep{oke68}.  If the spectral energy distribution (SED) is not flat, 
this will include both
the effect of detecting light from a region of the emitted spectrum shifted from the
effective wavelength of the detector and the effective squeezing of the detector
bandpass in the emitted frame. 

Following the formalism of \citet{hog02}, the {\it K} correction is defined as
\begin{equation}
M_{emitted} = m_{observed} - DM - K(z),
\end{equation}
where $DM$ is the distance modulus, given as
\begin{equation}
DM = 5\log_{10}\left[{D_L\over 10{\rm pc}}\right],
\end{equation}
$K(z)$ is the {\it K} correction, which depends on the SED of the observed
object \citep[see][eq. 8]{hog02}, and $D_L$ is the luminosity distance.  
Following \citet{hog00} and assuming $\Omega_M + \Omega_\Lambda=1$, 
the luminosity distance is given by
\begin{eqnarray}
d_L(z;\Omega_M,\Omega_\Lambda,H_0) = {c(1+z)\over H_0}\times\nonumber\\
\int_0^z[(1+z^{\prime})^2(1+\Omega_Mz^{\prime})
-z^{\prime}(2+z^{\prime})\Omega_\Lambda]^{-1/2}dz^{\prime}.
\end{eqnarray}
We assume $H_0 = $70km s$^{-1}$ Mpc$^{-1}$, $\Omega_M = 0.3$, and $\Omega_\lambda = 0.7$
throughout \citep{spe07}.

For a power-law SED of the form $f_\nu \propto \nu^\alpha$ we can write
\begin{equation}
K(z) = -2.5\alpha\log(1+z) - 2.5\log(1+z)
\end{equation}
where the first term transforms from the emitted band to the observed
band at z=0 and the second term corrects for the effective narrowing of the passband
when observing redshifted objects. Quasar spectra also contain various broad
emission lines, and their contribution to the {\it K} correction will be addressed
below. 

Table 2 (column 4) lists the absolute magnitude, $M_{i(z=0)}$, for each quasar as 
reported in the SDSS DR3Q catalog.  These values were computed
by correcting the SDSS $i$ magnitudes for Galactic extinction,  
assuming $H_0 = $70km s$^{-1}$ Mpc$^{-1}$, $\Omega_M = 0.3$, and $\Omega_\lambda = 0.7$,
and applying a standard continuum {\it K} correction by assuming a spectral index 
of $\alpha = -0.5$ and correcting to $z=0$.  

The DR3QLF is computed as a function of $M_i$, but
$K$ corrected to $z=2$, near the peak in the quasar distribution.  They also correct
for emission lines by convolving a composite spectrum created from 16,713 SDSS quasars
with the continuum subtracted with the SDSS filters.  Their values are reported in
Table 2 (column 5) as $M_{i(z=2)}$.  They compute a combined
{\it K} correction, assuming a spectral index of $\alpha = -0.5$ (see their Table 4),
which we will refer to here as $K_2$.
We adopt the \citet{ric06a} methodology but employ the spectral index computed for
our SQIID subsample in computing the $M_i$.  This gives an expression for $K(z)$:
\begin{equation}
K(z) = K_2 - 2.5(\alpha_o - \alpha_{fix})\log_{10}(1+z)
\end{equation} 
where $\alpha_{fix} = -0.5$.  Therefore $M_{i(\alpha_o)}$ is given as
\begin{equation}
M_{i(\alpha_o)} = m_i - 5\log_{10}\left[D_L\over 10{\rm pc}\right] \\
- K_2 + 2.5(\alpha_o + 0.5)\log_{10}(1+z).
\end{equation} 

The values of $\alpha_o$ and $M_{i(\alpha_o)}$ for the SQIID sample are listed in 
Table 2 (columns 6 and 8), and
the $M_{i(\alpha_o)}$ are plotted vs. redshift in Figure 5.  
Because our imaged $z\sim 4$ quasars are fainter than those in the two lower 
redshift ranges, the program samples quasars of similar luminosity at all 
three redshifts, unlike the 2MASS crossmatches, which have a distinct dependence
of luminosity on redshift, as would be expected for a flux limited
survey.  

\subsection{Correlations}

In order to investigate possible correlations of $\alpha_o$ with redshift 
or luminosity, we used the Astronomy SURVival Analysis tools 
\citep[ASURV;][]{lav92}, accessed through the {\textsc IRAF STSDAS} package.  
Linear regressions were performed using both the EM 
\citep[estimate and maximize;][]{dem77} 
and the Buckley-James \citep{bj79} algorithms
resulting in the following expressions for
$\alpha_o$ in terms of $z$ and $M_{i(\alpha_o)}$: 
\begin{equation}
\alpha_o(z) = (0.148\pm\ 0.068)z - (0.964\pm\ 0.200)
\end{equation}
\begin{equation}
\alpha_o(M_{i(\alpha_o)}) = (0.168\pm\ 0.083)M_{i(\alpha_o)} + (4.084\pm\ 2.305)
\end{equation}
\begin{equation}
\alpha_o(z,M_{i(\alpha_o)}) = (0.249\pm\ 0.065)z + (0.295\pm\ 0.080)M_{i(\alpha_o)} + (6.891\pm\ 2.137).
\end{equation}
The expressions for $\alpha_o(z)$ and $\alpha_o(M_{i(\alpha_o)})$ are plotted along
with the data in Figure 6.  We have also computed the generalized Kendall's tau
correlation coefficent between these variables using the BHK method \citep{lav92}.  
For the $\alpha_o$ and $z$ relation,
we find $\tau = 0.40$ with a 95\% probability of a correlation.  For $\alpha_o$ and 
$M_{i(\alpha_o)}$, we find $\tau = 0.30$ with an 85\% probability of a correlation.  
We note, however, that the SQIID sample was not chosen to sample a broad
range of absolute magnitude.
Instead fainter sources were necessarily chosen at higher redshifts because
fewer bright quasars are present at these very high redshifts in the SDSS sample
due to the relatively fewer numbers of quasars at these epochs.
We have also included an expression for $\alpha_o(z,M_{i(\alpha_o)})$ and show the projections of
the residuals of this fit along with the residuals from the expressions for 
$\alpha_o(z)$ and $\alpha_o(M_{i(\alpha_o)})$ in Figure 7.  
The residuals in the two-dimensional expressions (top panels) are similar to those for the
three-dimensional expression (lower panels).  

The distributions
of the spectral indices for the SQIID and 2MASS detected samples are shown in Figure 8 (left panel).  
Also shown are the separate distributions of the SQIID subsamples at $z\sim1.88$, $2.82$, and
$4.03$ (right panel).  
The $\alpha_o$ of the 2MASS detected SDSS quasars were computed using the same method
as for the SQIID sample, using the optical bands completely redward of Lyman-$\alpha$ and
the available 2MASS data from the DR3Q.   
Sample statistics are given in Table 3.
The median of the complete SQIID sample distribution is flatter
than the mean with a value of $-0.47$.  The mean is shifted to steeper values by the 
presence of a red tail to the distribution.  
The median of the 2MASS sample distribution is $-0.57$.
The mean is again shifted redward by a red tail to $\langle\alpha_o\rangle = -0.70\pm 0.53$.
The distributions of the two samples look very similar.  However, performance
of a Kolmogorov-Smirnov \citep[K-S;][]{bab96,nr92} test on the two data sets shows
only a 38\% probability that the two distributions are drawn from the same 
parent distribution.  The differences could be due to 
dependence of $\alpha$ on luminosity, as has been suggested by this work and others 
\citep[e.g.][]{ste06}, as the SQIID and 2MASS samples cover different areas of luminosity/redshift
space (Figure 5.)  If we consider just those quasars from the 2MASS sample with
$-28 < M_{i(\alpha_o)} < -26.5$ (where the bulk of the SQIID sample lies), then the mean of the 
distribution is $\langle\alpha_o\rangle = -0.54\pm 0.37$
and the median is -0.47, almost identical to the SQIID sample, even though the bulk of these
quasars are at redshifts between 0.5 and 2 (Table 3).  

As is clear from Figure 5, the 2MASS
detected sample has a strong correlation of luminosity with redshift.
While this is also true, to a lesser degree, for the SQIID sample, in order to select a 
statistically significant sample at high redshift and stay close
to our desired redshift ranges, we had to observe fainter candidates
at the higher redshift bins, as can be seen in Figure 2.  This resulted in
the SQIID subsamples having very similar mean luminosities of $M_{i(\alpha_o)}=-27.15$, $-27.89$,
and $-27.87$ at $z\sim 1.88$, $2.82$, and $4.03$, respectively.  It is therefore
not surprising that we see stronger evidence in favor of a correlation of $\alpha_o$
with $z$ than with luminosity, as our survey was designed to explore the relation with
redshift.  We do note, however, that if we limit the 2MASS detected sample to have similar
absolute magnitudes to our sample, then the statistics for $\alpha_o$ become very similar (Table 3), even
though the mean reshifts are $1.44$ and $2.81$.  What we conclude is that there is evidence
for evolution of $\alpha_o$ with both $z$ and luminosity, but that more data at a broader range
of luminosities is needed to fully characterize its form.  

Computing the difference in the absolute magnitudes in Table 2, one computed
with an average spectral index as in \citet{ric06a}, $M_{i(z=2)}$, and another
where the value of $\alpha_o$ is used, $M_{i(\alpha_o)}$, we can see in Figure
9 that there is a trend with redshift, due to the suggested evolution of $\alpha_o$
with redshift.  When using a spectral index computed from the quasar photometry, 
the quasars at $z\sim 2$ are brighter on average with respect to the luminosity computed with
$\alpha=-0.5$, while the quasars at $z\sim 4$ are fainter.  
The difference in computed luminosities in the SQIID sample
can be as much as $\pm1\fm8$. 

\subsection{The Quasar Luminosity Function}

If there is evolution in $\alpha$ with redshift or luminosity, this will have a direct effect
on the evolution of the QLF.  For example, the DR3QLF is shown in 
Figure 10 (open symbols).  If we recompute the absolute magnitudes of the bins 
\citep[see][Table 6]{ric06a} by using the mean redshift of the quasars in the bin in 
our Eq. 9 to compute an average $\alpha_o$ at this mean redshift, the absolute
magnitudes will shift by an amount $\Delta M_i = 2.5(\alpha_o + 0.5)\log_{10}(1+z)$.
This term will be zero where $\alpha_o = -0.5$, which corresponds to $z=3.1$.  For
redshifts significantly below this, the quasar SED is steeper on average, 
and $M_i$ is correspondingly brighter.  For redshifts above $3.1$, the SED's are
on average flatter, and the $M_i$ are fainter.  

This is shown graphically in Figure 10
where the DR3QLF is given along with points shifted in $M_i$ as described above.
While the space density remains essentially unchanged near the peak of quasar activity 
at $z\sim3$, the points shift towards brighter $M_i$ at lower redshift
and towards fainter $M_i$ at higher redshifts.  For cummulative space densities as a 
function of redshift, this would mean that at lower redshifts, more
objects would have $M_i$ brighter than a given cutoff, increasing space densities, while
at $z\sim4$, fewer objects would make the cutoff, resulting in lower space densities.  
This would change the shape of the form in the number of quasars over time, giving rise
to steeper growth at early times, with a more gradual decline locally.

\section{Discussion}
It has long been recognized that adopting an average
spectral index for the power-law form of quasar spectral energy distributions 
can effect the evolution in the QLF \citep[e.g.][]{gia92,fra96,ken97,wis98,ric06a}.
However, the adoption of an average spectral index persists in most studies of
the QLF, with $\alpha=-0.5$ being the most common value used.  Recent work
in the X-ray region has led to a general consensus that
there is a dependence of quasar SED on luminosity \citep[see][for
an exception]{tan07}. As for dependence on $z$, some groups find   
no evidence for the evolution of X-ray spectral indices, $\alpha_{ox}$,
with redshift \citep[e.g.][]{ste06}, while others do see a linear dependence of 
optical to X-ray spectral indices with $z$ \citep{kel07}.  X-ray emission
in AGN is generally taken to arise from a hot corona of optically thin gas heated by 
Compton scattering of thermal photons from a thick accretion disk,
the likely source of the optical/UV emission.  While emission from these two
regions is likely related, there is little evidence for a direct link between 
the X-ray and optical/UV emission, and \citet{kel07} find no evidence for a correlation
between $\alpha_{UV}$ and $\alpha_{ox}$.

Previous attempts to study the optical spectral energy distributions of quasars include
\citet{fra96} who find $\alpha=-0.46\pm0.30$ from a sample of LBQS quasars using 
optical and NIR photometry, \citet{cri90} who compute {\it K} corrections as a function 
of redshift in $UBV$ and find $\alpha\approx-0.7$ at $1000-5500\AA$ from their composite
quasar spectra, \cite{van01} who 
construct a composite quasar spectrum from an SDSS quasar sample and find $\alpha=-0.46$ 
for the region Ly-$\alpha$ to H$\beta$, and \citet{pen03} who find 
$\langle\alpha\rangle = -0.57 \pm 0.33$
from optical and NIR photometry of 45 $z > 3.60$ SDSS quasars.  
There is considerable spread in
the values of $\alpha$ within each sample, and the distribution of $\alpha$
found here and by the groups using optical and NIR photometry \citep{fra96,pen03} are remarkably
similar, each having a peak around -0.5 to -0.3 with a tail to the red that shifts the mean
of the samples to steeper values.  This could be consistent with the findings of \citet{web95} in their
comparison of optical to NIR colors of radio quiet and radio loud quasars, who predict quasars
should have an SED with $\alpha=-0.3$ and that the distribution is caused by dust reddening in the
host galaxies.  
Perhaps more important than the evolution of $\alpha_o$ with $z$ is the distribution of the
spectral index values.  \citet{laf97} have pointed out that, if there is a spread in the
spectral slope, there will be a corresponding slower luminosity evolution
and steeper QLF's.    

In general, the most desirable way to present the QLF is in terms of the
bolometric luminosity.  
\citet{ric06b} demonstrate that computing bolometric luminosities from optical 
luminosities assuming a single mean quasar SED can lead to errors as large as 50\%. 
\citet{hop07} have determined the bolometric QLF by combining the results
of over two dozen quasar surveys from the hard X-ray to the mid-IR.  They construct
a model SED but allow for the distribution in the power-law components of the model,
stressing that there is no ``effective mean'' SED.
They also adopt the luminosity dependent value of $\alpha_{ox}$ of
\citet{ste06}.  However, they do not adopt a value of $\alpha_{ox}$ that depends
on redshift as has been suggested by \citet{kel07} and is supported by our findings
in the restframe optical. 
 
Given the brightness of our sample, we have not attempted to correct for the contribution of the
host galaxy to the SED.  The optical and NIR data were taken several years apart, and we have
not considered variability in this sample.  
We have obtained NIR observations for $\sim100$ more SDSS quasars in these redshift
ranges, along with nearly simultaneous ($\sim 1$ month) optical imaging with the aim of 
adding to our survey sample and addressing the issue of variability.
\citet{ric03} have stressed the need to correct for redshift dependent color effects
when computing photometric spectral indexes, as failing to do so can lead to effects
systematic with redshift.  
We have neglected the presence of 
emission lines in our photometric passbands.  However, 
due to the nature of the sample - each quasar is sampled at essentially the same
points in their SED's as is clearly seen in Figure 5 -
the apparent evolution would persist.

The generally accepted view of quasar activity is ascribed to the release of gravitational
energy by accretion of matter on to 
a supermassive black hole.  The UV/optical flux is seen as arising from a thin, optically
thick accretion disk \citep[see][for a review]{kor99}, so
a correlation between $\alpha_{o}$ and $z$ implies evolution in the accretion process.
More studies of the form and possible evolution of quasar SED's are obviously still needed to
constrain theoretical models of AGN structure and energy production mechanisms.  
While ever larger samples of quasars are discovered \citep[e.g.][]{sch07}, 
the form of the QLF cannot be fully characterized until we have
a better understanding of quasar energy distributions and how they are affected by 
luminosity, redshift, and environment.  Since we are now coming to understand how quasar
activity might be related to galaxy formation and evolution, quantifying the shape of the
QLF and its evolution remains a pressing problem.  Here we have presented some evidence that
quasar SED's evolve with cosmic time and have shown that this has a direct effect on
the evolution of their luminosity function.

\acknowledgments
J.K. and S.B acknowledge the support of NSF ADVANCE grant AST-0340837 and 
NVO Research Initiative Grant \#000012.  We thank the staff at KPNO for
their assistance at the 2.1m telescope, especially Mike Merrill for his
assistance with the SQIID instrument and the {\textsc UPSQIID} package.  
We also thank our anonymous referee whose suggestions greatly improved
the manuscript.

This research has made use of data obtained from and software provided by the 
US National Virtual Observatory, which is sponsored by the National Science Foundation.

This publication makes use of data products from the Two Micron All Sky Survey, 
which is a joint project of the University of Massachusetts and the Infrared 
Processing and Analysis Center/California Institute of Technology, funded by the 
National Aeronautics and Space Administration and the National Science Foundation.

Funding for the creation and distribution of the SDSS Archive has been provided 
by the Alfred P. Sloan Foundation, the Participating Institutions, the National 
Aeronautics and Space Administration, the National Science Foundation, the U.S. 
Department of Energy, the Japanese Monbukagakusho, and the Max Planck Society. 
The SDSS Web site is http://www.sdss.org/.
The Participating Institutions are The University of Chicago, Fermilab, the 
Institute for Advanced Study, the Japan Participation Group, The Johns Hopkins 
University, the Max-Planck-Institute for Astronomy (MPIA), the Max-Planck-Institute 
for Astrophysics (MPA), New Mexico State University, Princeton University, the 
United States Naval Observatory, and the University of Washington.

{\it Facilities:} \facility{KPNO:2.1m(SQIID)}, \facility{NVO}, \facility{Sloan}.

\begin{figure}
\epsscale{1.1}
\plotone{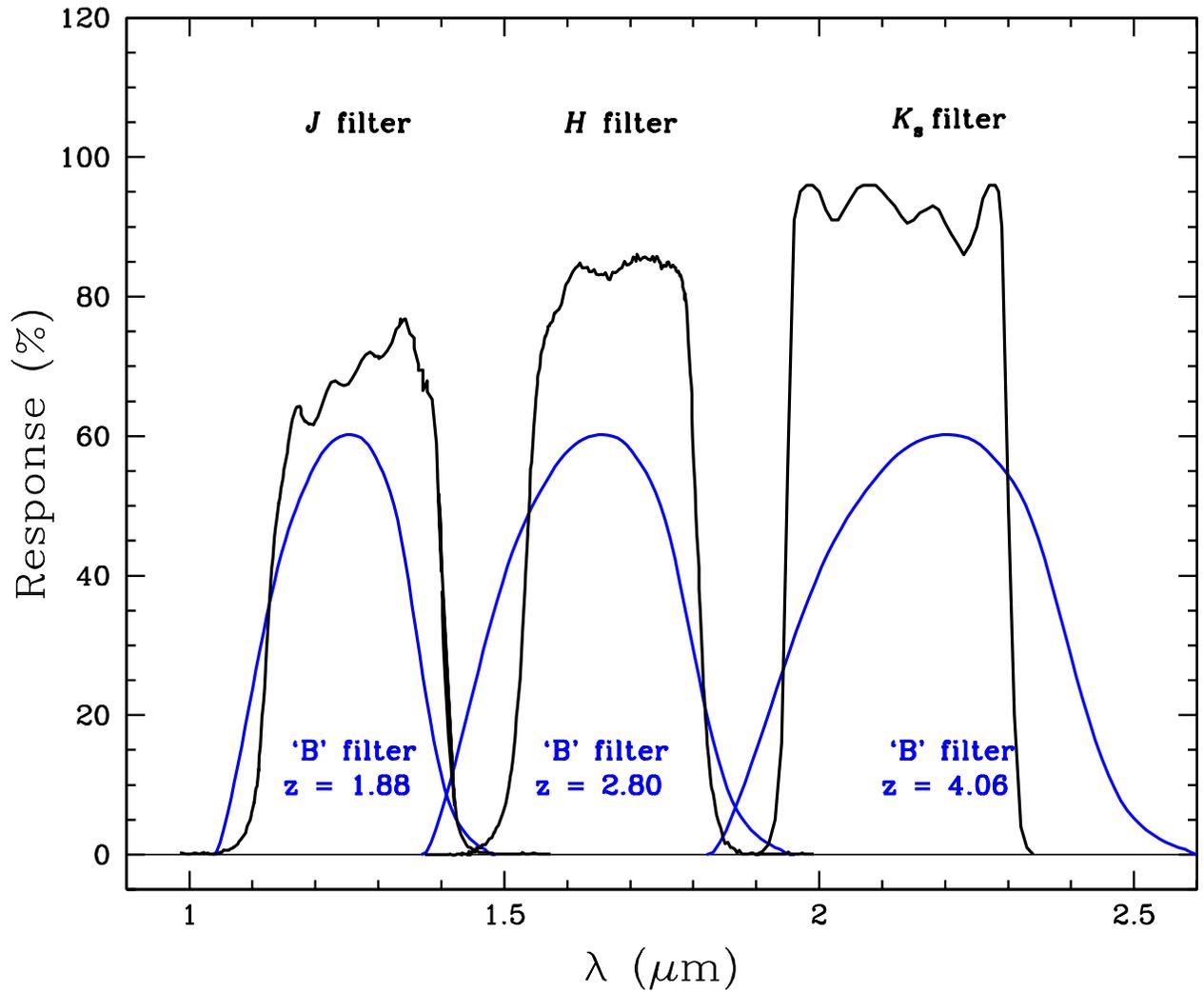}
\caption{SQIID $JHK_s$ filters along with the Johnson $B$ filter ``redshifted''
to show where the restframe $B$-band flux of quasars at the given redshift 
would be with respect to the corresponding SQIID NIR filter.}
\end{figure}

\begin{figure}
\epsscale{1.1}
\plotone{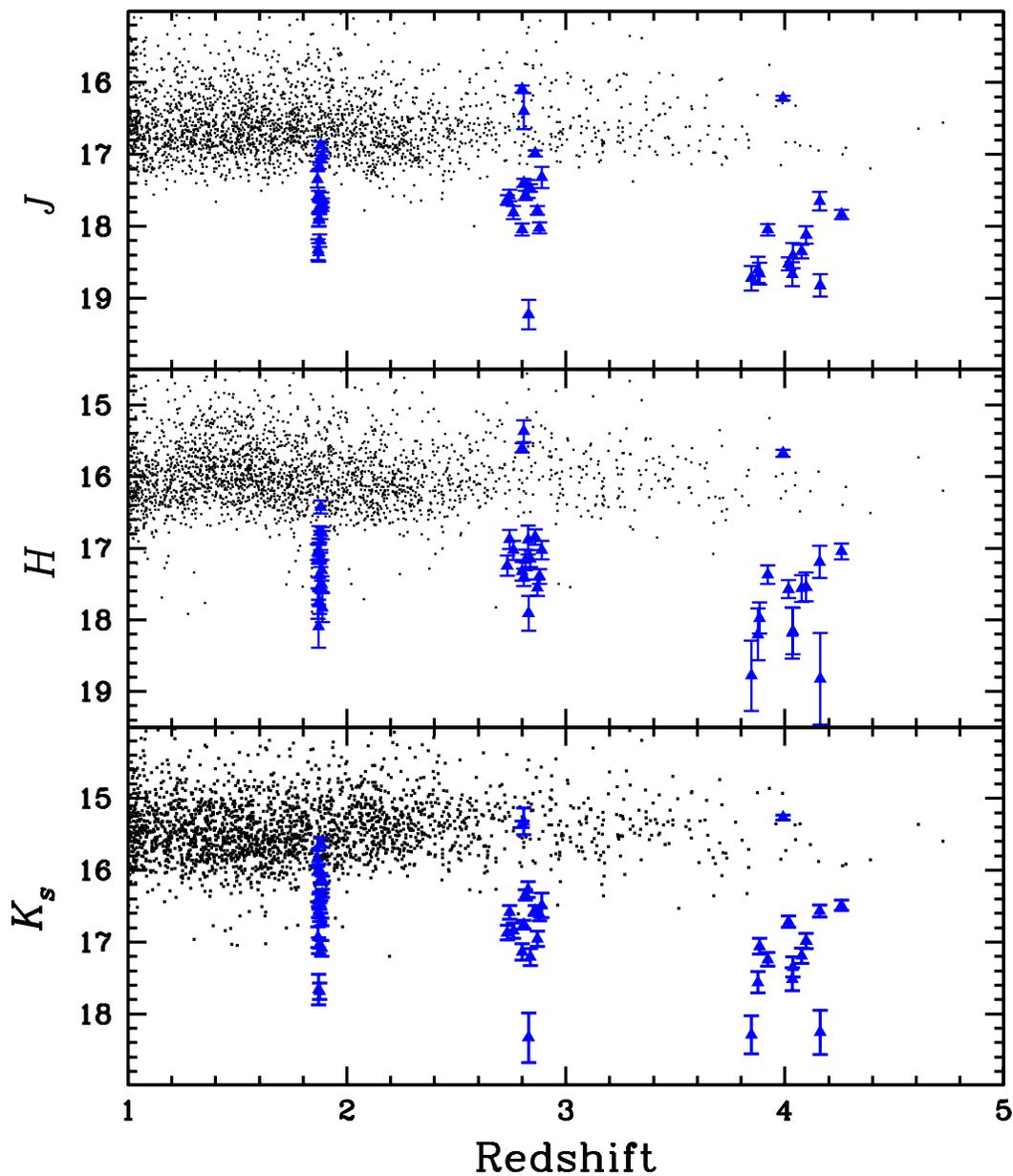}
\caption{SQIID $JHK_s$ magnitudes for the sub-sample of SDSS quasars given 
in Table 1 (blue triangles) along with their associated errors.  Also shown 
are the $JHK_s$ magnitudes of the 6192 quasars from the SDSS Third Data 
Release Quasar Catalog \citep{sch05} with measurable 2MASS detections 
(black dots).  The SQIID data reach about $1^m$ to $2^m$  fainter than the 
2MASS data.}
\end{figure}

\begin{figure}
\epsscale{1.1}
\plotone{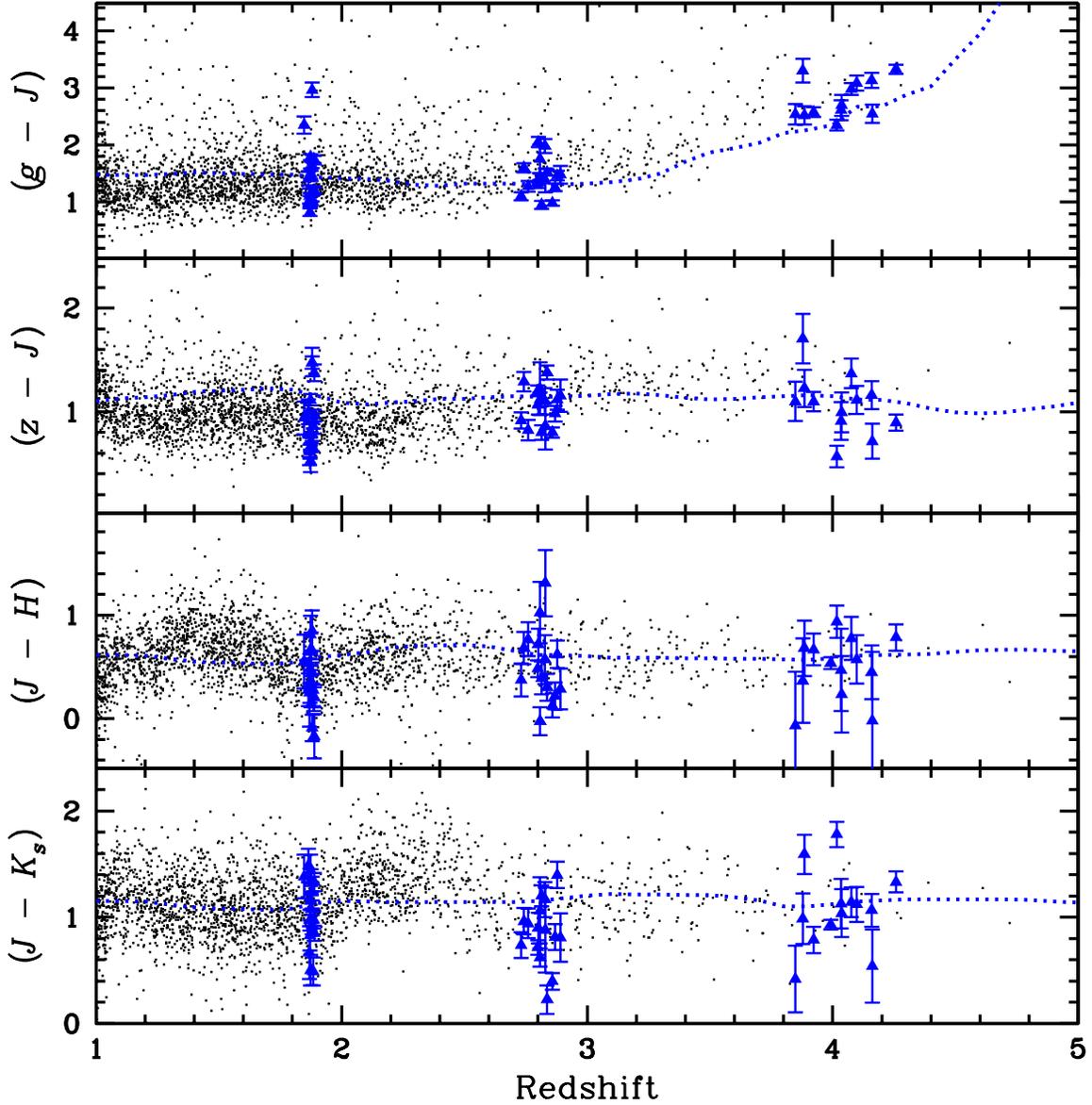}
\caption{SDSS optical and SQIID $JHK_s$ colors for the sub-sample of SDSS 
quasars observed with SQIID given 
in Table 1 (blue triangles) along with their associated errors.  Also shown 
are the colors of the 6192 quasars from the SDSS Third Data 
Release Quasar Catalog with measurable 2MASS detections 
(black dots). The dashed line shows the expected colors of quasars
with spectral indices of $\alpha = -0.5$, including emission lines and 
Lyman-$\alpha$ forest continuum depression.  
}
\end{figure}

\begin{figure}
\epsscale{1.1}
\plotone{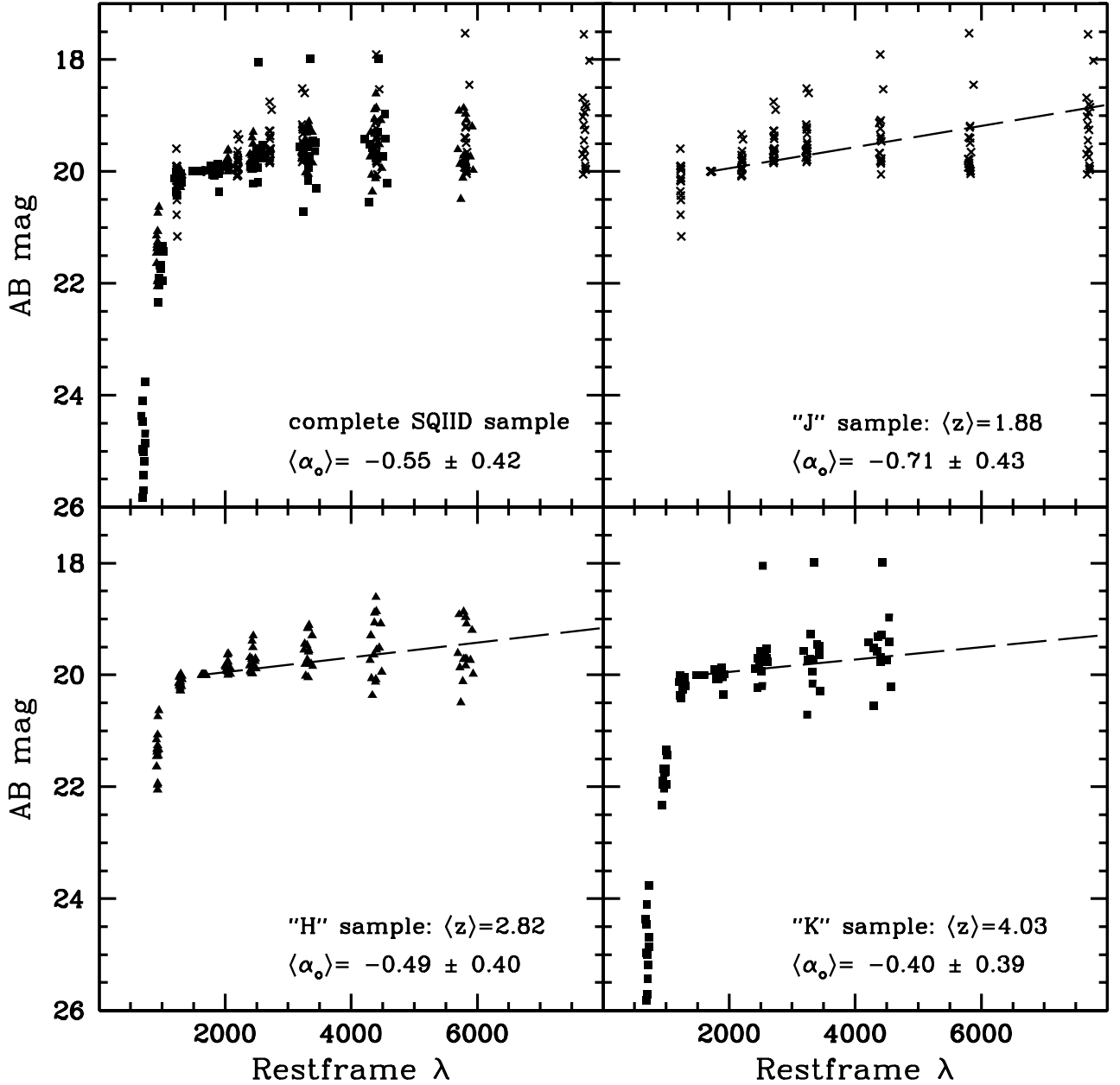}
\caption{
AB magnitude vs. restframe wavelength for the SQIID sample.  The $JHK$ magnitudes
have been converted to the AB system (see text; the SDSS magnitudes are based
on the AB system), and all magnitudes shifted to the restframe of the quasar.  The
magnitudes are also normalized to 20.0 in the bluest band completely redward of
Lyman-$\alpha$, $g$ for the $\langle z\rangle=1.88$ sample, $r$ for the
$\langle z\rangle=2.82$ sample, and $i$ for the $\langle z\rangle=4.03$ sample.
A power law is fit to each unnormalized quasar SED separately and
given in Table 2.  The broken lines show the SED of a power law spectrum
with a mean $\alpha_o$ for the sample.
}
\end{figure}

\begin{figure}
\epsscale{1.1}
\plotone{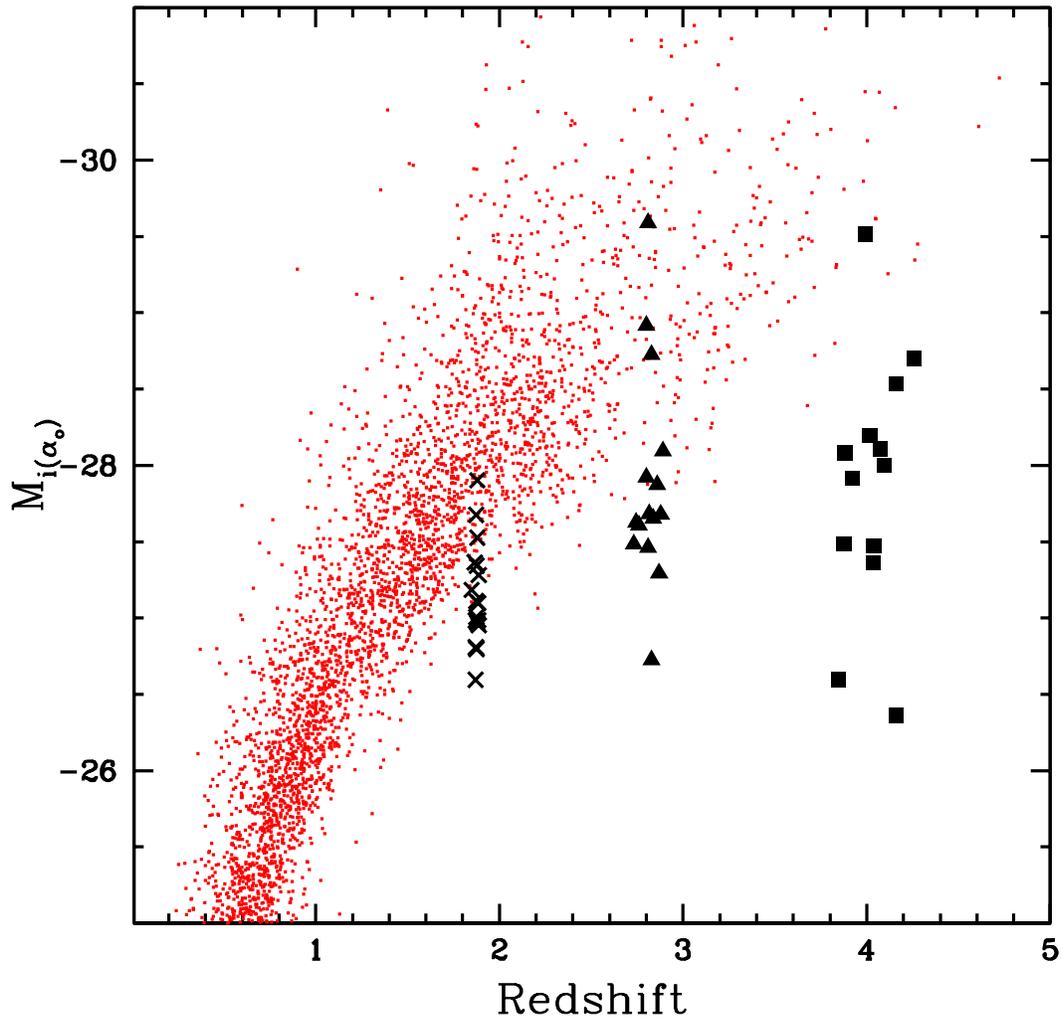}
\caption{Absolute magnitude $M_{i(\alpha_o)}$ vs. redshift for the SQIID sample (filled black symbols).
Crosses represent quasars at $z\sim1.88$, 
triangles at $z\sim2.82$ and squares at $z\sim4.03$.  The red dots respresent SDSS
quasars with 2MASS detections.
}
\end{figure}

\begin{figure}
\epsscale{1.1}
\plottwo{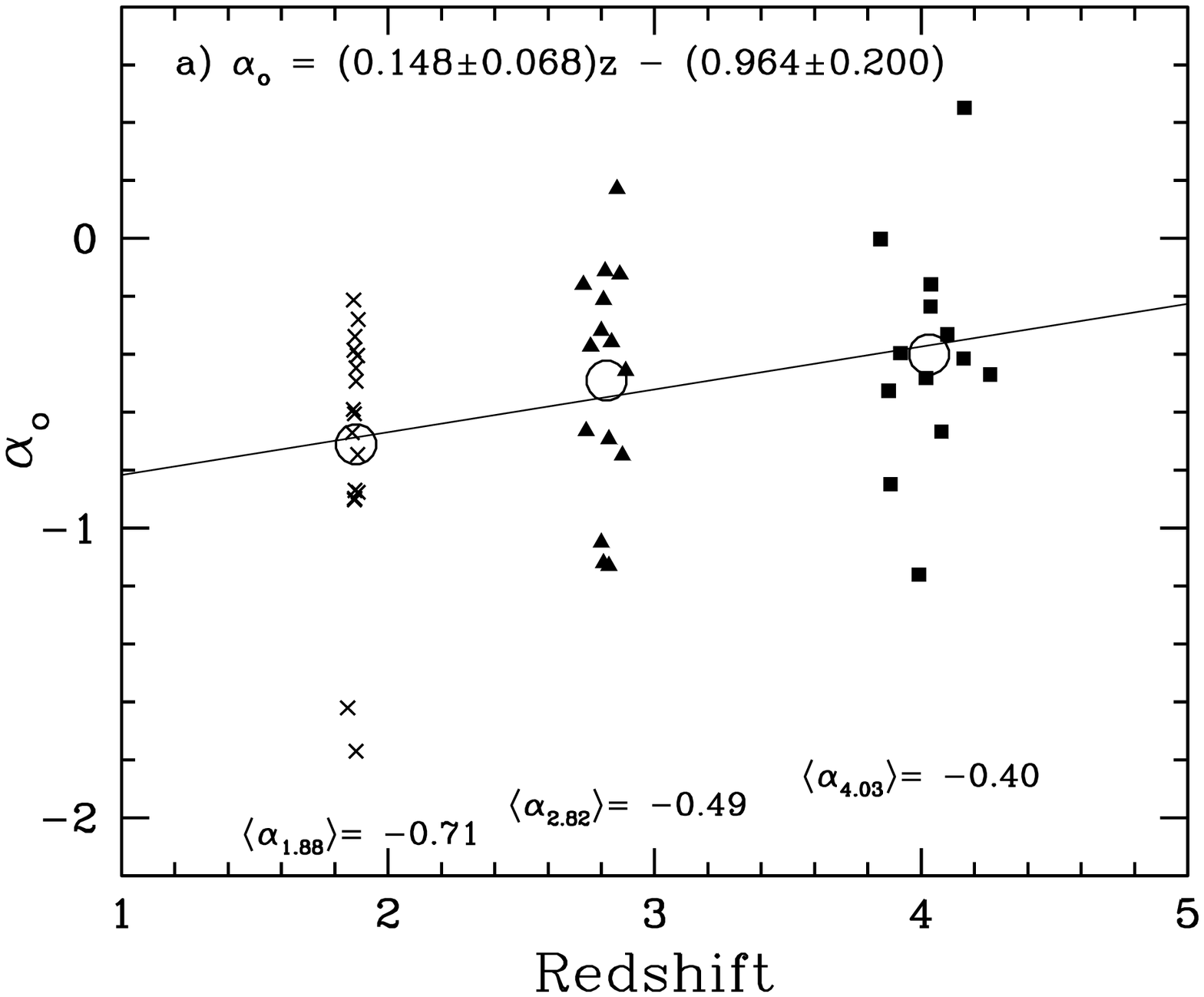}{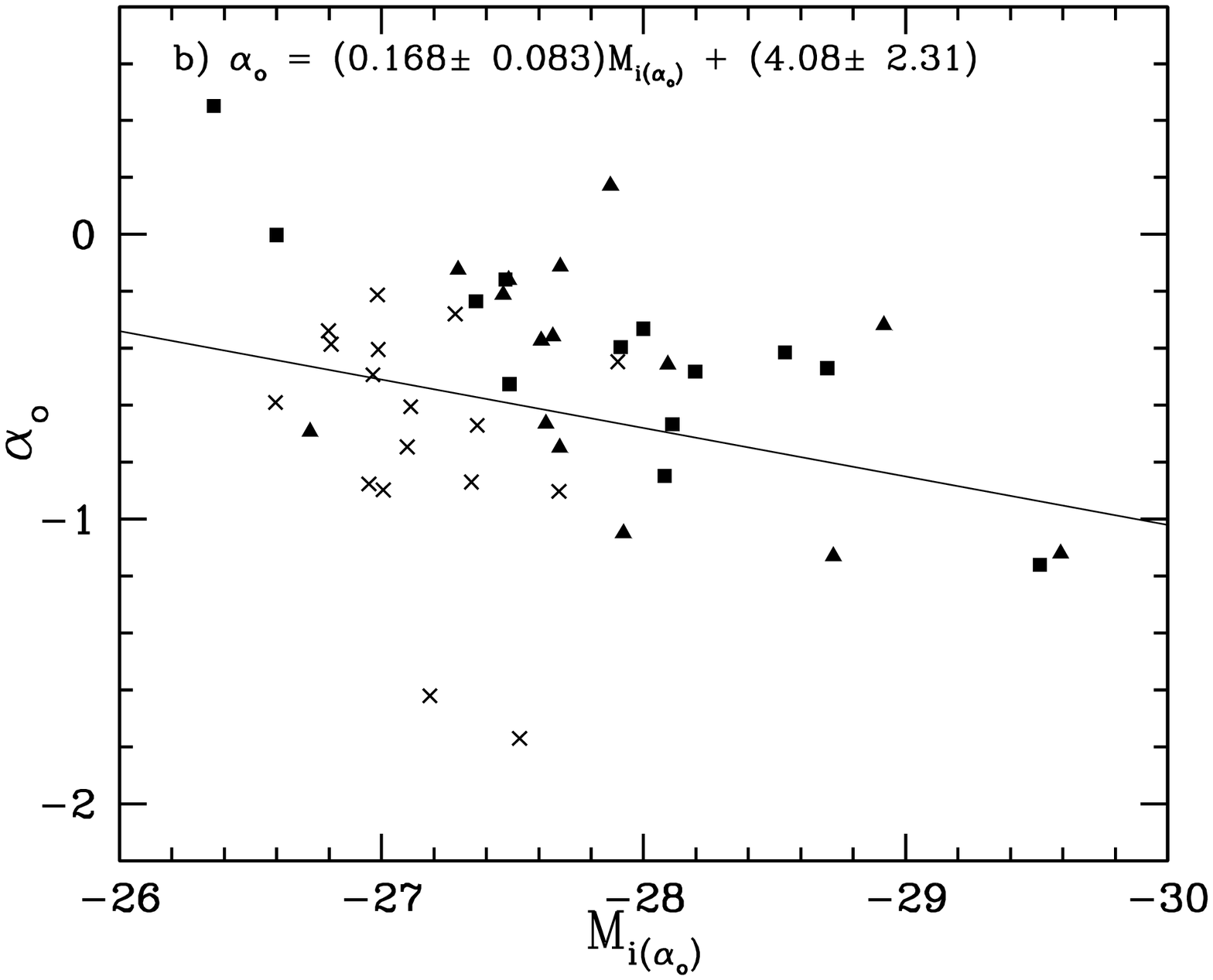}
\caption{a) Spectral index $\alpha_o$ vs. redshift for the SDSS subsample observed
with SQIID (closed symbols.)  Also shown is the mean $\alpha_o$ for each redshift
bin (open circles) with the value given below.  The line is a linear fit to the
sample given by the relation given at upper left.  The mean for the
sample as a whole is $\langle\alpha_o\rangle = -0.55 \pm 0.42$.
b)  Spectral index $\alpha_o$ vs. $M_{i(\alpha_o)}$ for the SQIID sample.  
The quasars at $z\sim 1.88$, $2.82$,
and $4.03$ are shown as crosses, triangles, and squares, respectively.  The fit to
the data as a function of $M_{i(\alpha_o)}$ is given in the upper left and shown as a solid line.      
}
\end{figure}

\begin{figure}
\epsscale{0.9}
\plotone{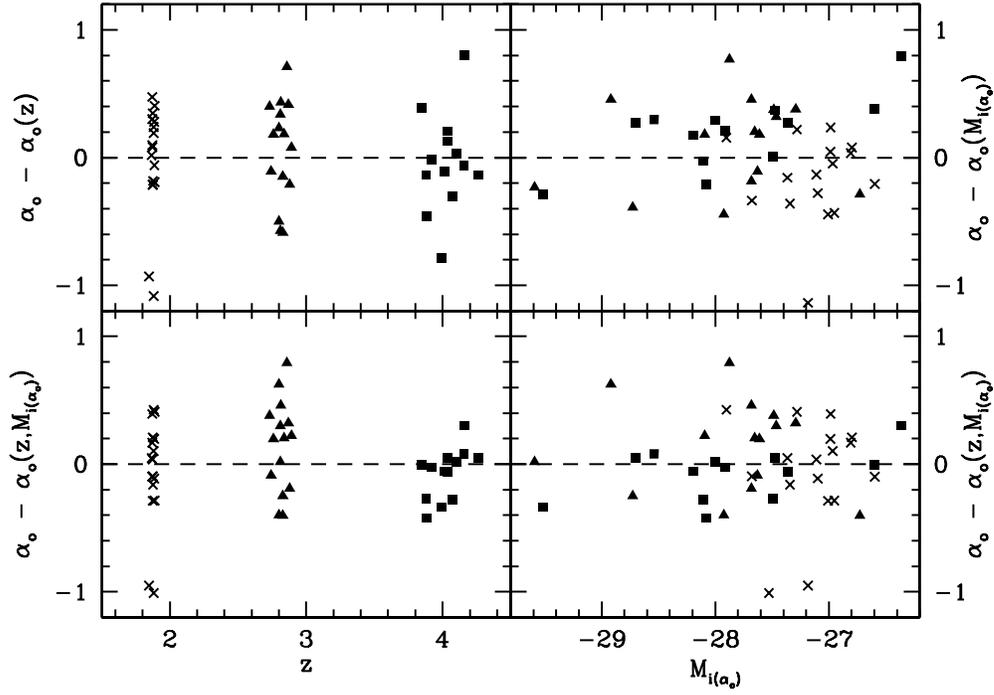}
\caption{
(Top panels) Residuals in $\alpha_o$ of the linear expressions given in Equations 9 and 10.  
(Bottom panels) Two dimensional projections of the residuals in $\alpha_o$ of the expression
given in Equation 11.  The residuals average to $\sim0$, but exhibit
strong scatter, with residuals as large as $\sim1$ in all cases.  
}
\end{figure}

\begin{figure}
\epsscale{1.1}
\plotone{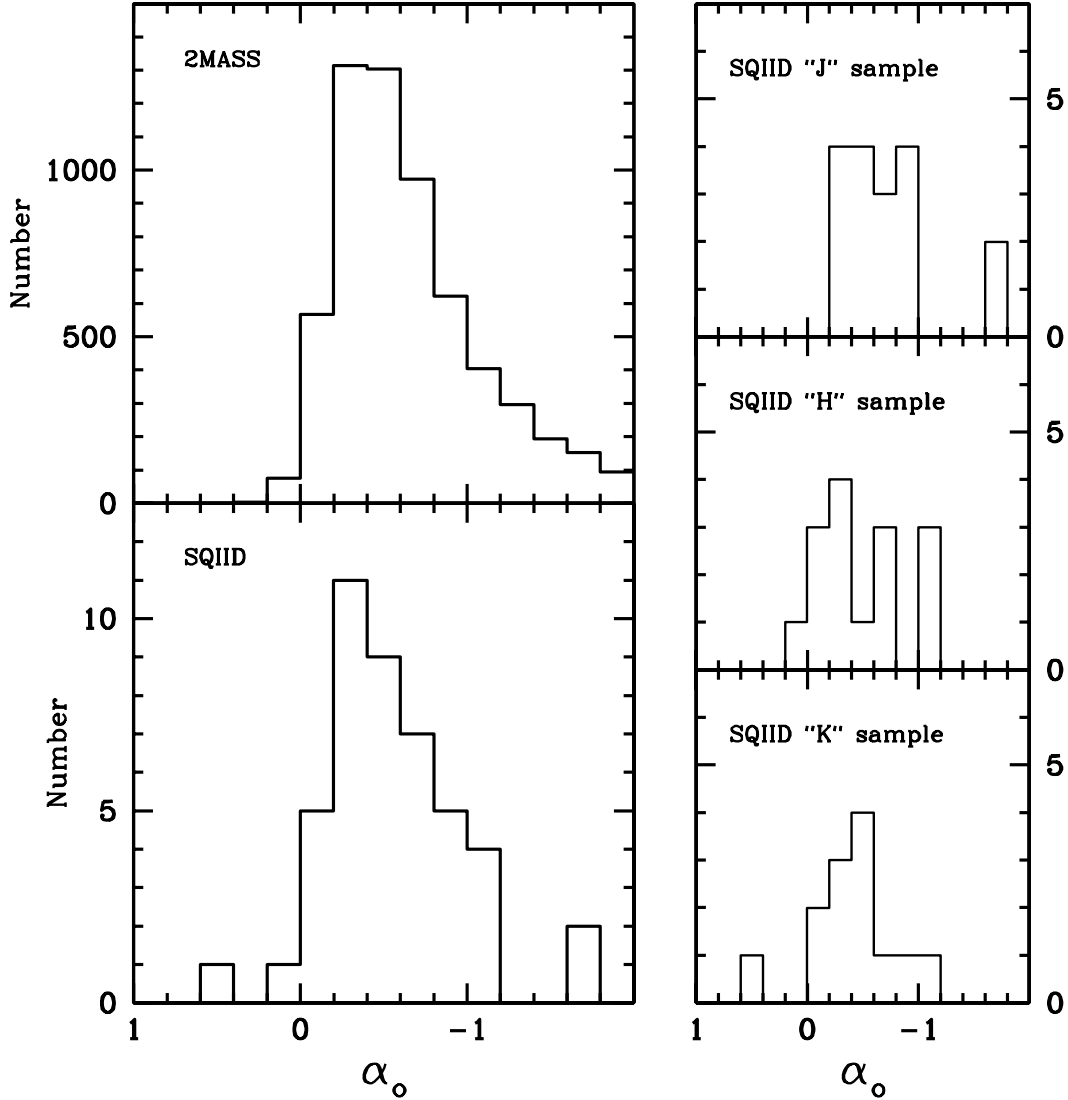}
\caption{Left: (top) The distribution of restframe optical spectral indices, $\alpha_o$, for
the SDSS quasars detected by 2MASS.  
The mean of the sample is $\langle\alpha\rangle = -0.70\pm 0.53$
and the median is $-0.57$. 
(bottom) The distribution of $\alpha_o$ for the SQIID sample. The mean of the sample is 
$\langle\alpha\rangle = -0.55\pm 0.42$ and the median is $-0.47$.
Right: The distribution of $\alpha_o$ for the SQIID sample in each redshift bin.  
See Table 3 for sample statistics.  
}
\end{figure}

\begin{figure}
\epsscale{1.1}
\plottwo{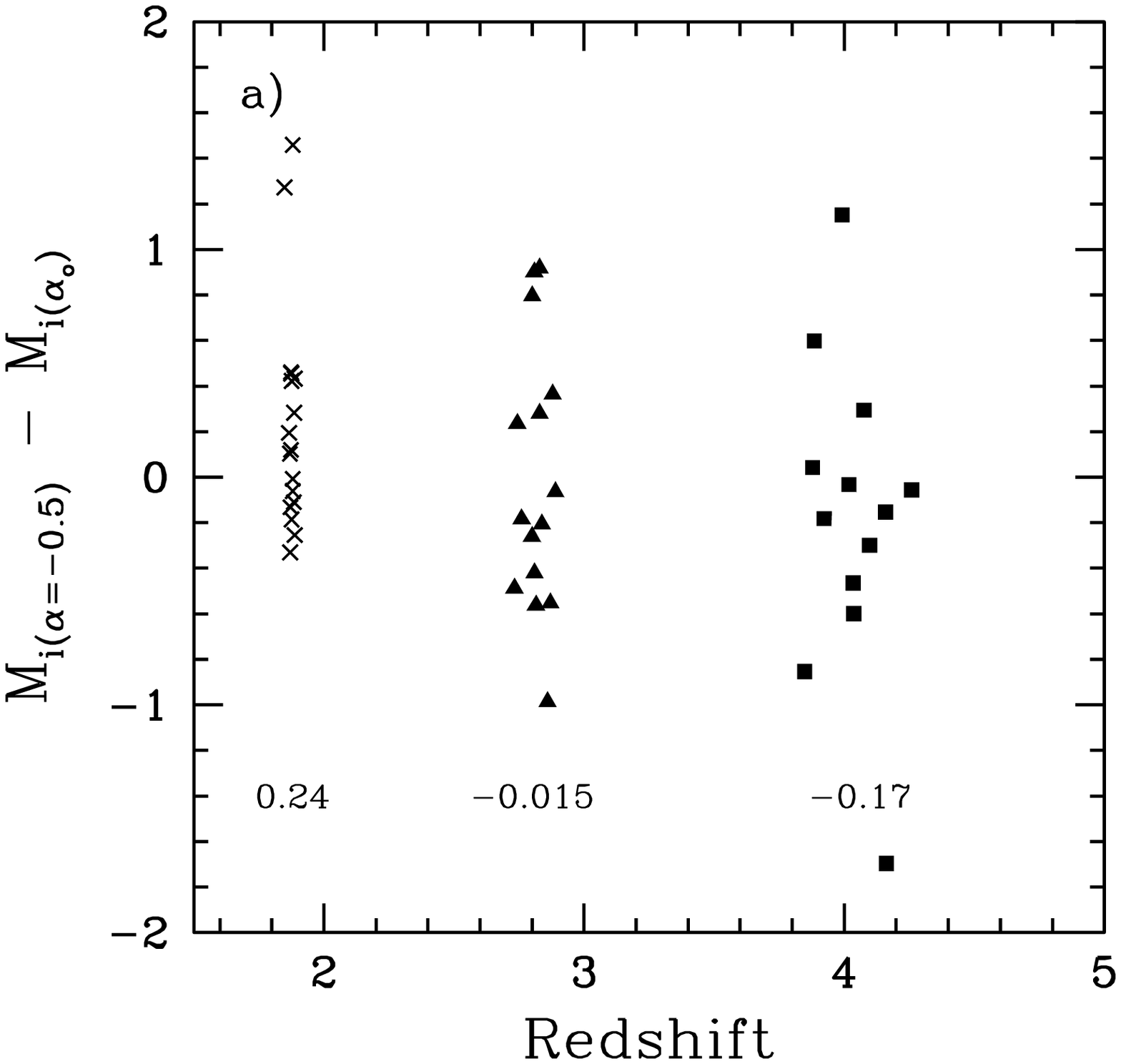}{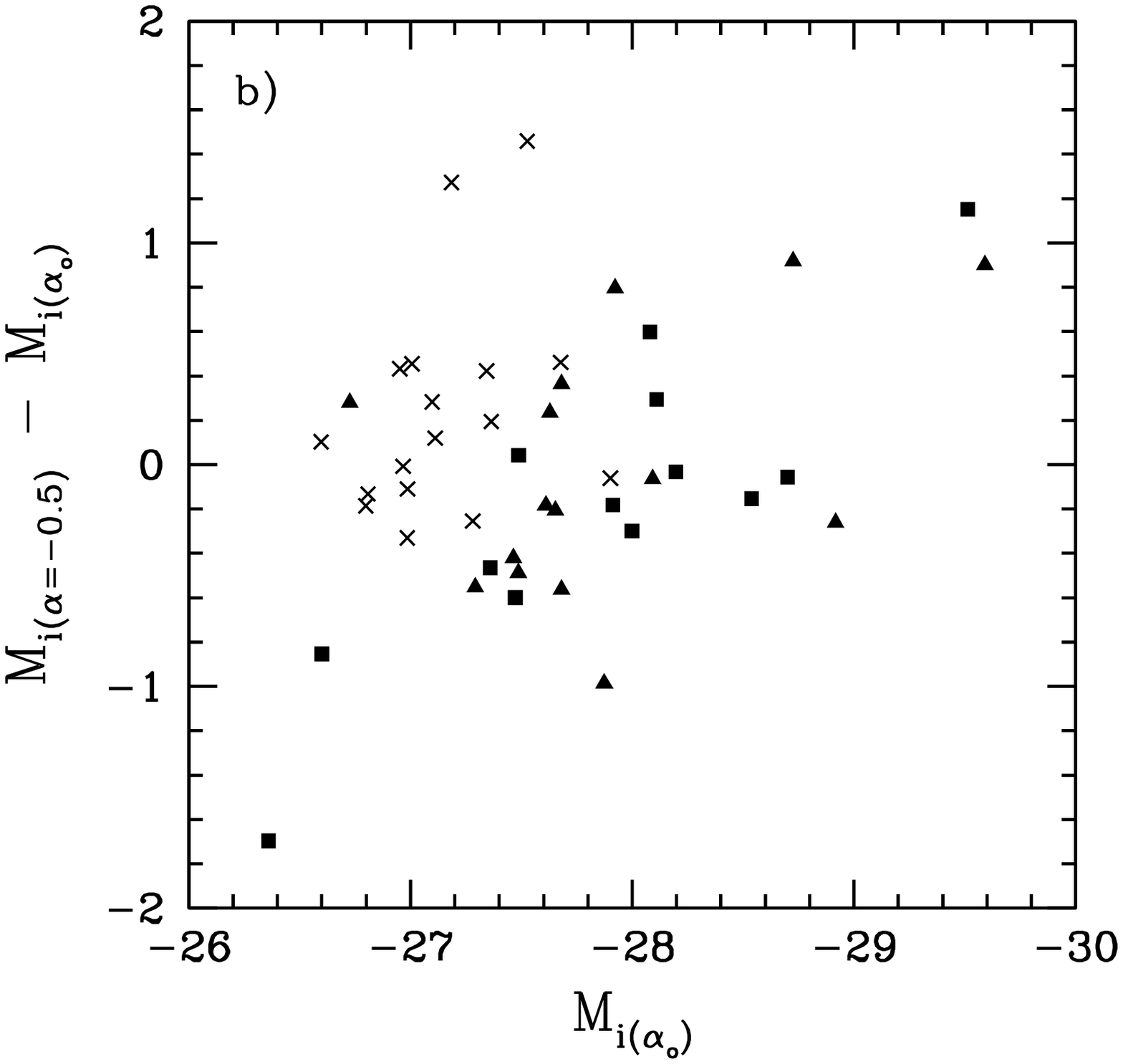}
\caption{a) The difference in absolute magnitude, $M_i$, assuming $\alpha=-0.5$ and
the $\alpha_o$ from Table 2, as a function of redshift, for the SQIID sample.
Also given are the average difference in magnitude for each redshift interval.
There is a trend with redshift reflecting the trend toward flatter contiuum
slopes at higher redshifts for the sample, as demonstrated in Figure 6a.
b) The same difference as a function of $M_i(\alpha_o)$.  
The quasars at $z\sim 1.88$, $2.82$,
and $4.03$ are shown as crosses, triangles, and squares, respectively.
}
\end{figure}

\begin{figure}
\epsscale{1.1}
\plotone{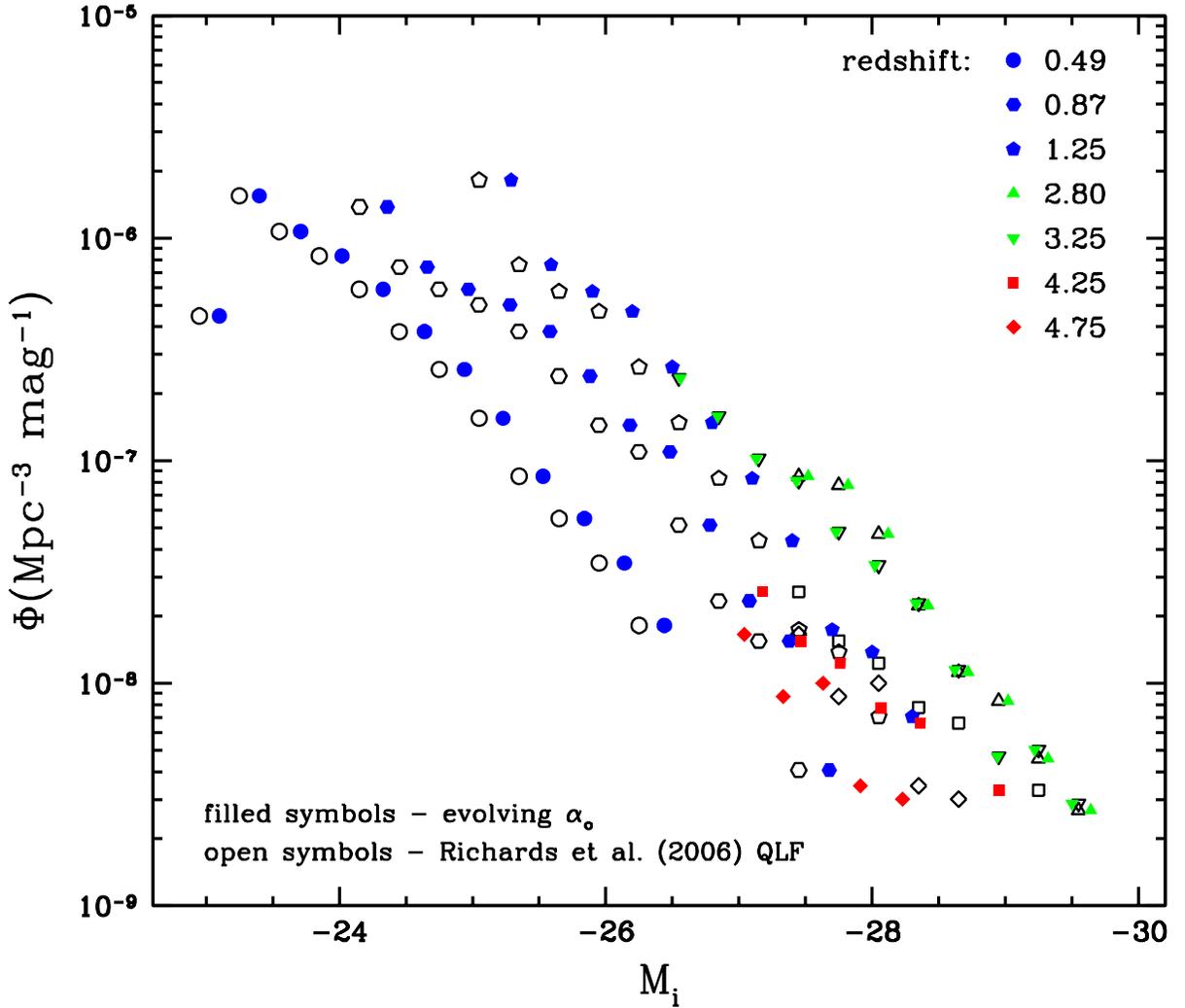}
\caption{The QLF of \citet{ric06a} (black open symbols) and the same QLF
but with the $M_i$ of the redshift bins calculated using an evolving $\alpha$ as
given by Equation 9 and using the mean redshift of the quasars in the
bin (colored filled symbols).  Lower redshift quasars are found to be
brighter and high redshift quasars are found to be fainter.  This results
in a steeper form for the evolution in quasar space densities.  
}
\end{figure}

\clearpage

\begin{deluxetable}{lccccc}
\tablewidth{0pt}
\tabletypesize{\scriptsize}
\tablecolumns{6}
\tablecaption{SQIID Sample Quasar Near-IR Photometry\label{tbl-1}}
\tablehead{
\colhead{SDSS DR3 designation} & \colhead{Redshift} & 
\colhead{$i^*$} &
\colhead{$J$} &
\colhead{$H$} &
\colhead{$K_s$} 
}
\startdata
 SDSS 095048.48$-$000017.7  &  1.8802  &  19.562$\pm$0.024  &  17.778$\pm$0.121  &  16.909$\pm$0.153  &  16.513$\pm$0.089 \\
 SDSS 095938.28$-$003500.8  &  1.8753  &  18.593$\pm$0.018  &  17.728$\pm$0.100  &  17.374$\pm$0.210  &  16.630$\pm$0.085 \\
 SDSS 101119.94$-$004145.3  &  1.8879  &  19.098$\pm$0.020  &  17.676$\pm$0.064  &  17.831$\pm$0.201  &  16.323$\pm$0.058 \\
 SDSS 102517.58$+$003422.0  &  1.8879  &  18.091$\pm$0.014  &  17.011$\pm$0.051  &  16.787$\pm$0.083  &  16.134$\pm$0.051 \\
 SDSS 103204.74$-$001119.1  &  1.8715  &  18.690$\pm$0.021  &  17.915$\pm$0.089  &  17.563$\pm$0.156  &  17.052$\pm$0.111 \\
 SDSS 103427.57$-$002233.9  &  1.8698  &  19.153$\pm$0.024  &  18.366$\pm$0.128  &  18.089$\pm$0.302  &  17.659$\pm$0.214 \\
 SDSS 110725.70$+$003353.8  &  1.8732  &  18.581$\pm$0.018  &  17.896$\pm$0.103  &  17.071$\pm$0.147  &  16.434$\pm$0.102 \\
 SDSS 115115.38$+$003826.9  &  1.8805  &  17.593$\pm$0.016  &  16.869$\pm$0.047  &  16.423$\pm$0.091  &  15.617$\pm$0.066 \\
 SDSS 121655.39$+$001415.3  &  1.8706  &  18.233$\pm$0.014  &  17.570$\pm$0.061  &  17.079$\pm$0.127  &  16.581$\pm$0.075 \\
 SDSS 123505.91$-$003022.3  &  1.8804  &  18.584$\pm$0.016  &  17.062$\pm$0.048  &  17.144$\pm$0.115  &  16.120$\pm$0.079 \\
 SDSS 123514.94$+$004740.7  &  1.8747  &  19.000$\pm$0.018  &  18.199$\pm$0.087  &  17.755$\pm$0.159  &  17.684$\pm$0.115 \\
 SDSS 123947.61$+$002516.2  &  1.8483  &  19.585$\pm$0.035  &  18.326$\pm$0.147  &  17.770$\pm$0.213  &  16.934$\pm$0.142 \\
 SDSS 132742.92$+$003532.6  &  1.8736  &  18.337$\pm$0.016  &  17.155$\pm$0.049  &  16.778$\pm$0.089  &  15.986$\pm$0.071 \\
 SDSS 135605.41$-$010024.4  &  1.8860  &  18.812$\pm$0.015  &  17.722$\pm$0.063  &  17.504$\pm$0.111  &  16.720$\pm$0.051 \\
 SDSS 141015.36$-$001418.9  &  1.8758  &  18.673$\pm$0.017  &  17.743$\pm$0.087  &  17.055$\pm$0.133  &  16.379$\pm$0.074 \\
 SDSS 143641.24$+$001558.9  &  1.8659  &  18.401$\pm$0.015  &  17.347$\pm$0.114  &  17.055$\pm$0.109  &  15.839$\pm$0.110 \\
 SDSS 145838.04$+$002417.9  &  1.8847  &  18.520$\pm$0.015  &  17.602$\pm$0.069  &  17.309$\pm$0.142  &  17.086$\pm$0.110 \\
\\[-5pt]
 SDSS 094745.26$-$004113.2  &  2.8287  &  18.922$\pm$0.017  &  17.480$\pm$0.124  &  16.883$\pm$0.202  &  16.270$\pm$0.109 \\
 SDSS 100423.27$-$004042.9  &  2.7320  &  18.627$\pm$0.014  &  17.637$\pm$0.067  &  17.245$\pm$0.142  &  16.869$\pm$0.103 \\
 SDSS 102832.09$-$004607.0  &  2.8592  &  17.839$\pm$0.017  &  16.986$\pm$0.041  &  16.842$\pm$0.109  &  16.560$\pm$0.068 \\
 SDSS 105808.47$+$003930.5  &  2.8149  &  18.392$\pm$0.023  &  17.568$\pm$0.065  &  17.161$\pm$0.145  &  16.342$\pm$0.073 \\
 SDSS 121323.94$+$010414.7  &  2.8292  &  20.185$\pm$0.038  &  19.229$\pm$0.203  &  17.910$\pm$0.245  &  18.329$\pm$0.345 \\
 SDSS 121920.26$+$010736.1  &  2.8005  &  19.453$\pm$0.022  &  18.047$\pm$0.083  &  17.323$\pm$0.128  &  17.137$\pm$0.116 \\
 SDSS 121933.25$+$003226.4  &  2.8791  &  19.342$\pm$0.030  &  18.023$\pm$0.079  &  17.394$\pm$0.104  &  16.611$\pm$0.095 \\
 SDSS 122730.37$-$010446.1  &  2.8701  &  18.798$\pm$0.026  &  17.780$\pm$0.063  &  17.554$\pm$0.114  &  16.954$\pm$0.105 \\
 SDSS 124551.44$+$010505.0  &  2.8088  &  17.896$\pm$0.014  &  16.397$\pm$0.252  &  15.368$\pm$0.153  &  15.321$\pm$0.184 \\
 SDSS 125241.55$-$002040.6  &  2.8909  &  18.520$\pm$0.013  &  17.320$\pm$0.150  &  17.023$\pm$0.129  &  16.492$\pm$0.171 \\
 SDSS 131128.35$+$004929.7  &  2.8090  &  18.726$\pm$0.015  &  17.395$\pm$0.059  &  17.408$\pm$0.120  &  16.757$\pm$0.062 \\
 SDSS 133647.14$-$004857.1  &  2.7997  &  17.413$\pm$0.020  &  16.091$\pm$0.049  &  15.600$\pm$0.064  &  15.360$\pm$0.047 \\
 SDSS 143307.40$+$003319.0  &  2.7432  &  19.176$\pm$0.022  &  17.571$\pm$0.082  &  16.877$\pm$0.131  &  16.587$\pm$0.094 \\
 SDSS 145754.03$+$003639.0  &  2.7603  &  18.820$\pm$0.019  &  17.810$\pm$0.092  &  17.027$\pm$0.134  &  16.840$\pm$0.105 \\
 SDSS 150611.23$+$001823.6  &  2.8377  &  18.832$\pm$0.016  &  17.466$\pm$0.053  &  17.145$\pm$0.119  &  17.209$\pm$0.121 \\
\\[-5pt]
 SDSS 094822.96$+$005554.4  &  3.8777  &  20.083$\pm$0.037  &  18.605$\pm$0.183  &  18.203$\pm$0.362  &  17.559$\pm$0.150 \\
 SDSS 104837.40$-$002813.6  &  3.9918  &  19.094$\pm$0.068  &  16.219$\pm$0.030  &  15.672$\pm$0.049  &  15.265$\pm$0.033 \\
 SDSS 105254.59$-$000625.8  &  4.1619  &  19.535$\pm$0.024  &  18.823$\pm$0.153  &  18.824$\pm$0.643  &  18.255$\pm$0.309 \\
 SDSS 105602.37$+$003222.0  &  4.0361  &  19.411$\pm$0.028  &  18.409$\pm$0.176  &  18.160$\pm$0.324  &  17.349$\pm$0.140 \\
 SDSS 105902.73$+$010404.0  &  4.0978  &  19.208$\pm$0.021  &  18.122$\pm$0.123  &  17.541$\pm$0.200  &  16.984$\pm$0.108 \\
 SDSS 110813.85$-$005944.5  &  4.0175  &  19.254$\pm$0.020  &  18.523$\pm$0.091  &  17.571$\pm$0.126  &  16.717$\pm$0.077 \\
 SDSS 111224.18$+$004630.3  &  4.0346  &  19.647$\pm$0.026  &  18.667$\pm$0.170  &  18.184$\pm$0.361  &  17.519$\pm$0.160 \\
 SDSS 120138.56$+$010336.2  &  3.8475  &  19.867$\pm$0.028  &  18.723$\pm$0.172  &  18.781$\pm$0.491  &  18.291$\pm$0.263 \\
 SDSS 121531.55$-$004900.4  &  3.8842  &  19.863$\pm$0.037  &  18.660$\pm$0.150  &  17.975$\pm$0.219  &  17.058$\pm$0.107 \\
 SDSS 122600.68$+$005923.6  &  4.2586  &  18.859$\pm$0.021  &  17.836$\pm$0.064  &  17.044$\pm$0.111  &  16.490$\pm$0.073 \\
 SDSS 131052.50$-$005533.2  &  4.1585  &  18.847$\pm$0.023  &  17.650$\pm$0.129  &  17.193$\pm$0.225  &  16.569$\pm$0.084 \\
 SDSS 135828.74$+$005811.3  &  3.9225  &  19.318$\pm$0.022  &  18.048$\pm$0.079  &  17.368$\pm$0.132  &  17.240$\pm$0.096 \\
 SDSS 141315.36$+$000032.3  &  4.0760  &  19.700$\pm$0.027  &  18.352$\pm$0.092  &  17.562$\pm$0.186  &  17.188$\pm$0.106 \\
\enddata

\tablecomments{Table \ref{tbl-1} Optical $i^*$ photometry and redshift data are taken from the SDSS DR3Q \citep{sch05}.}

\end{deluxetable}


\begin{deluxetable}{lccccccc}
\tablewidth{0pt}
\tabletypesize{\scriptsize}
\tablecolumns{8}
\tablecaption{SQIID Sample Spectral Indices\label{tbl-2}}
\tablehead{
\colhead{SDSS DR3 designation} & 
\colhead{Redshift} & 
\colhead{$A_u$} &
\colhead{$M_{i(z=0)}$\tablenotemark{a}} &
\colhead{$M_{i(z=2)}$\tablenotemark{b}} &
\colhead{$\alpha_o$} & 
\colhead{$\sigma_{\alpha_o}$} & 
\colhead{$M_{i(\alpha_o)}$\tablenotemark{c}}
}
\startdata
 SDSS 095048.48$-$000017.7  &  1.8802  &  0.298  &  -25.777  & -26.068  &  -1.77  &  0.12 &  -27.527 \\
 SDSS 095938.28$-$003500.8  &  1.8753  &  0.183  &  -26.694  & -26.983  &  -0.34  &  0.09 &  -26.798 \\
 SDSS 101119.94$-$004145.3  &  1.8879  &  0.234  &  -26.225  & -26.519  &  -0.88  &  0.07 &  -26.952 \\
 SDSS 102517.58$+$003422.0  &  1.8879  &  0.257  &  -27.241  & -27.535  &  -0.28  &  0.05 &  -27.281 \\
 SDSS 103204.74$-$001119.1  &  1.8715  &  0.330  &  -26.651  & -26.940  &  -0.39  &  0.07 &  -26.808 \\
 SDSS 103427.57$-$002233.9  &  1.8698  &  0.375  &  -26.205  & -26.493  &  -0.59  &  0.10 &  -26.596 \\
 SDSS 110725.70$+$003353.8  &  1.8732  &  0.181  &  -26.702  & -26.990  &  -0.61  &  0.07 &  -27.111 \\
 SDSS 115115.38$+$003826.9  &  1.8805  &  0.114  &  -27.672  & -27.963  &  -0.45  &  0.04 &  -27.902 \\
 SDSS 121655.39$+$001415.3  &  1.8706  &  0.130  &  -27.026  & -27.315  &  -0.21  &  0.06 &  -26.985 \\
 SDSS 123505.91$-$003022.3  &  1.8804  &  0.121  &  -26.684  & -26.975  &  -0.49  &  0.05 &  -26.967 \\
 SDSS 123514.94$+$004740.7  &  1.8747  &  0.126  &  -26.263  & -26.552  &  -0.90  &  0.08 &  -27.007 \\
 SDSS 123947.60$+$002516.2  &  1.8483  &  0.087  &  -25.629  & -25.912  &  -1.62  &  0.14 &  -27.185 \\
 SDSS 132742.92$+$003532.6  &  1.8736  &  0.134  &  -26.928  & -27.216  &  -0.90  &  0.05 &  -27.677 \\
 SDSS 135605.41$-$010024.4  &  1.8860  &  0.264  &  -26.521  & -26.815  &  -0.75  &  0.06 &  -27.098 \\
 SDSS 141015.36$-$001418.9  &  1.8758  &  0.222  &  -26.630  & -26.920  &  -0.87  &  0.07 &  -27.343 \\
 SDSS 143641.24$+$001558.9  &  1.8659  &  0.208  &  -26.884  & -27.171  &  -0.67  &  0.06 &  -27.365 \\
 SDSS 145838.04$+$002417.9  &  1.8847  &  0.251  &  -26.806  & -27.098  &  -0.40  &  0.06 &  -26.987 \\
\\[-5pt]
 SDSS 094745.26$-$004113.2  &  2.8287  &  0.416  &  -27.390  & -27.807  &  -1.13  &  0.09 &  -28.725 \\
 SDSS 100423.27$-$004042.9  &  2.7320  &  0.281  &  -27.553  & -27.974  &  -0.16  &  0.07 &  -27.486 \\
 SDSS 102832.09$-$004607.0  &  2.8592  &  0.276  &  -28.440  & -28.859  &   0.17  &  0.06 &  -27.874 \\
 SDSS 105808.47$+$003930.5  &  2.8149  &  0.216  &  -27.828  & -28.245  &  -0.11  &  0.06 &  -27.682 \\
 SDSS 121323.94$+$010414.7  &  2.8292  &  0.171  &  -26.028  & -26.445  &  -0.69  &  0.17 &  -26.727 \\
 SDSS 121920.26$+$010736.1  &  2.8005  &  0.102  &  -26.710  & -27.126  &  -1.05  &  0.11 &  -27.923 \\
 SDSS 121933.25$+$003226.4  &  2.8791  &  0.130  &  -26.893  & -27.314  &  -0.75  &  0.09 &  -27.681 \\
 SDSS 122730.37$-$010446.1  &  2.8701  &  0.121  &  -27.427  & -27.846  &  -0.12  &  0.09 &  -27.293 \\
 SDSS 124551.44$+$010505.0  &  2.8088  &  0.103  &  -28.274  & -28.690  &  -1.12  &  0.07 &  -29.591 \\
 SDSS 125241.55$-$002040.6  &  2.8909  &  0.159  &  -27.736  & -28.158  &  -0.46  &  0.08 &  -28.093 \\
 SDSS 131128.35$+$004929.7  &  2.8090  &  0.161  &  -27.467  & -27.883  &  -0.21  &  0.06 &  -27.464 \\
 SDSS 133647.14$-$004857.1  &  2.7997  &  0.140  &  -28.764  & -29.181  &  -0.32  &  0.04 &  -28.918 \\
 SDSS 143307.40$+$003319.0  &  2.7432  &  0.186  &  -26.975  & -27.393  &  -0.66  &  0.08 &  -27.628 \\
 SDSS 145754.03$+$003639.0  &  2.7603  &  0.266  &  -27.377  & -27.794  &  -0.37  &  0.08 &  -27.610 \\
 SDSS 150611.23$+$001823.6  &  2.8377  &  0.309  &  -27.444  & -27.861  &  -0.36  &  0.08 &  -27.654 \\
\\[-5pt]
 SDSS 094822.96$+$005554.4  &  3.8777  &  0.562  &  -26.977  & -27.445  &  -0.53  &  0.22 &  -27.488 \\
 SDSS 104837.40$-$002813.6  &  3.9918  &  0.208  &  -27.885  & -28.361  &  -1.16  &  0.06 &  -29.513 \\
 SDSS 105254.59$-$000625.8  &  4.1619  &  0.254  &  -27.552  & -28.056  &   0.45  &  0.18 &  -26.360 \\
 SDSS 105602.37$+$003222.0  &  4.0361  &  0.212  &  -27.593  & -28.073  &  -0.16  &  0.17 &  -27.473 \\
 SDSS 105902.73$+$010404.0  &  4.0978  &  0.157  &  -27.806  & -28.298  &  -0.33  &  0.13 &  -27.999 \\
 SDSS 110813.85$-$005944.5  &  4.0175  &  0.243  &  -27.753  & -28.230  &  -0.48  &  0.10 &  -28.197 \\
 SDSS 111224.18$+$004630.3  &  4.0346  &  0.186  &  -27.346  & -27.825  &  -0.24  &  0.16 &  -27.360 \\
 SDSS 120138.56$+$010336.2  &  3.8475  &  0.111  &  -26.993  & -27.454  &  -0.00  &  0.19 &  -26.600 \\
 SDSS 121531.55$-$004900.4  &  3.8842  &  0.096  &  -27.012  & -27.480  &  -0.85  &  0.15 &  -28.080 \\
 SDSS 122600.68$+$005923.6  &  4.2586  &  0.123  &  -28.223  & -28.757  &  -0.47  &  0.11 &  -28.701 \\
 SDSS 131052.50$-$005533.2  &  4.1585  &  0.131  &  -28.188  & -28.692  &  -0.41  &  0.11 &  -28.539 \\
 SDSS 135828.74$+$005811.3  &  3.9225  &  0.190  &  -27.616  & -28.095  &  -0.40  &  0.10 &  -27.913 \\
 SDSS 141315.36$+$000032.3  &  4.0760  &  0.219  &  -27.328  & -27.816  &  -0.67  &  0.14 &  -28.110 \\
\enddata

\tablecomments{Table \ref{tbl-2} Redshift and galactic extinction data are taken from the SDSS DR3Q \citep{sch05}.}
\tablenotetext{a}{Absolute $M_i$ from SDSS DR3Q, column 32, $K$-corrected to $z=0$}
\tablenotetext{b}{Absolute $M_i$ from \citet{ric06a} who $K$-correct to $z=2$}
\tablenotetext{c}{Absolute $M_i$ computed using values from \citet{ric06a} who
$K$-correct to $z=2$, but using the spectral index, $\alpha_o$, in column 6.}

\end{deluxetable}


\begin{deluxetable}{lcccccc}
\tablewidth{0pt}
\tabletypesize{\scriptsize}
\tablecolumns{6}
\tablecaption{Sample Spectral Index Statistics}
\tablehead{
\colhead{} & 
\colhead{} & 
\colhead{} &
\colhead{} &
\colhead{} &
\colhead{$\alpha_o$} &
\colhead{}\\  
\cline{5-7}
\colhead{Quasar Sample} & 
\colhead{N} & 
\colhead{$\langle z \rangle$} &
\colhead{$\langle M_{i(\alpha_o)} \rangle$} &
\colhead{median} &
\colhead{mean} &
\colhead{$\sigma$}  
}
\startdata
SDSS w/SQIID				&  45	&  2.81   & -27.61 & -0.47 &	-0.55 &	0.42 \\
~~~``J'' subsample			&  17	&  1.88	  & -27.15 & -0.61 &	-0.71 &	0.43 \\
~~~``H'' subsample			&  15	&  2.82	  & -27.89 & -0.37 &	-0.49 &	0.40 \\
~~~``K'' subsample			&  13	&  4.03	  & -27.87 & -0.41 &	-0.40 &	0.39 \\
\\[-5pt]
SDSS w/2MASS				&  6192	&  0.98	  & -25.62 & -0.57 &	-0.70 &	0.53 \\
SDSS w/2MASS ($-28 < M_i < -26.5$)	&  1336	&  1.44	  & -27.29 & -0.47 &	-0.54 &	0.37 \\
\enddata

\end{deluxetable}



\clearpage
\begin{thebibliography}{}

\bibitem[Babu \& Feigelson (1996)]{bab96} Babu, G.J. \& Feigelson, E.D. 1996, 
{\it Astrostatistics: Interdisciplinary Statistics} (1st Edition, London: Chapman \& Hall)

\bibitem[Bonnarel et al.\ (2000)]{bon00} Bonnarel, F., Fernique, P., Bienaym\'{e}, O., 
	Egret, D., Genova, F., Louys, M., Ochsenbein, F., Wenger, M., \& 
	Bartlett, J.G. 2000, \aaps, 143, 33

\bibitem[Boyle et al.\ (1987)]{boy87} Boyle, B.J., Fong, R., Shanks, T., \& Peterson,
	B.A. 1987, \mnras, 227, 717

\bibitem[Buckley \& James (1979)]{bj79} Buckley, J., \& James, I. 1979, Biometrika, 66, 429

\bibitem[Cristiani \& Vio (1990)]{cri90} Cristiani, S. \& Vio, R. 1990, \aap, 227, 385

\bibitem[Croom et al.\ (2004)]{cro04} Croom, S.M., Smith, R.J., Boyle, B.J., Shanks, T.,
	Miller, L., Outram, P.J., \& Loaring, N.S. 2004, \mnras, 349, 1397

\bibitem[Cutri et al.\ (2003)]{cut03} Cutri, R.M. et al.\ 2003, {\it The IRSA 2MASS All-Sky
	Point Source Catalog}, http://irsa.ipac.caltech.edu/applications/Gator/

\bibitem[Dempster et al.\ (1977)]{dem77} Dempster, A.P., Laird, N.M., \& Rubin, D.B. 1977, 
Royal Stat. Soc. B, 39, 1

\bibitem[Francis (1996)]{fra96} Francis, P.J. 1996, {\it Publ. Astron. Soc. Aust.}, 13, 212

\bibitem[Giallongo \& Vagnetti (1992)]{gia92} Giallongo, E. \& Vagnetti, F. 1992, \apj, 396, 411

\bibitem[Glikman et al.\ (2006)]{Gli06} Glikman, E., Helfand, D.J. \& White, R.L. 2006, 
	\apj, 640, 579

\bibitem[Hogg (2000)]{hog00} Hogg, D.W. 2000, preprint (astro-ph/9905116)

\bibitem[Hogg et al.\ (2002)]{hog02} Hogg, D.W., Baldry, I.K., Blanton, M.R., \& Eisenstein,
	D.J. 2002, preprint (astro-ph/0210394)

\bibitem[Holberg \& Bergeron (2006)]{hb06} Holberg, J.B. \& Bergeron, P. 2006, \aj, 
	132, 1221

\bibitem[Hopkins et al.\ (2007)]{hop07} Hopkins, P.F., Richards, G.T., \& Hernquist, L.
	2007, \apj, 654, 731

\bibitem [Humason, Mayall, \& Sandage (1956)]{hum56} Humason, M.L., Mayall, N.U., \& Sandage,
	A.R. 1956, \aj, 61, 97

\bibitem[Kelly et al.\ (2007)]{kel07} Kelly, B.C., Bechtold, J., Siemiginowska, A., Aldcroft, T., 
	\& Sobolewska, M. 2007, \apj, 657, 116

\bibitem[Kennefick et al.\ (1995)]{ken95} Kennefick, J.D., Djorgovski, S.G., \& de Carvalho,
	R.R. 1995, \aj, 110, 2553

\bibitem[Kennefick et al.\ (1997)]{ken97} Kennefick, J.D., Osmer, P.S., Pahre, M. 
	\& Djorgovski, S. 1997, in N.R. Tanvir, A. Aragon-Salamanca and J.V. Wall (eds.), 
	{\it The Hubble Space Telescope and the High Redshift Universe}, 
	Singapore: World Scientific Publishing Company, p. 401

\bibitem[Koratkar \& Blaes (1999)]{kor99} Koratkar, A.,, \& Blaes, O. 1999, \pasp, 111, 1

\bibitem[La Franca \& Cristiani (1997)]{laf97} La Franca, F. \& Cristiani, S. 1997, \aj, 113, 1517

\bibitem[LaValley et al.\ (1992)]{lav92} LaValley, M., Isobe, T., \& Feigelson, E. 1992,
A.S.P. Conference Series, 25, 245

\bibitem[M\'{e}gessier (1995)]{meg95} M\'{e}gessier, C. 1995, \aap, 296, 771

\bibitem[Oke \& Gunn (1983)]{og83} Oke, J.B. \& Gunn, J.E. 1983, \apj, 266, 713

\bibitem[Oke \& Sandage (1968)]{oke68} Oke, J.B. \& Sandage, A. 1968, \apj, 154, 21 

\bibitem[Pentericci et al.\ (2003)]{pen03} Pentericci, L., Rix, H.W., Prada, F., 
    	Fan, X., Strauss, M.A., Schneider, D.P., Grebel, E.K., Harbeck, D., Brinkmann, J., 
    	\& Narayanan, V.K. 2003, \aap, 410, 75

\bibitem[Press et al.\ (1992)]{nr92} Press, W.H, Teukolsky, S.A., Vetterling, W.T., \&
	Flannery, B.P. 1992, {\it Numerical Recipes in Fortran:  The Art of Scientific
	Computing} (2nd Edition, Cambridge: Cambridge University Press)

\bibitem[Richards et al.\ (2006a)]{ric06a} Richards, G.T. et al.\ 2006, \aj, 131, 2766

\bibitem[Richards et al.\ (2006b)]{ric06b} Richards, G.T. et al.\ 2006, \apjs, 166, 470

\bibitem[Richards et al.\ (2003)]{ric03} Richards, G.T. et al.\ 2003, \aj, 126, 1131

\bibitem[Schlegel et al.\ (1998)]{sch98} Schlegel, D.J., Finkbeiner, D.P., \& Davis, M. 
	1998, \apj, 500, 525

\bibitem[Schmidt \& Green (1983)]{sch83} Schmidt, M. \& Green, R.F. 1983, \apj, 269, 352

\bibitem[Schmidt, Schneider \& Gunn (1995)]{sch95} Schmidt, M., Schneider, D.P., \& Gunn, 
	J.E. 1995, \aj, 110, 68 

\bibitem[Schneider et al.\ (2002)]{sch02} Schneider, D.P. et al.\ 2002, \aj, 123, 567

\bibitem[Schneider et al.\ (2005)]{sch05} Schneider, D.P. et al.\ 2005, \aj, 130, 367

\bibitem[Schneider et al.\ (2007)]{sch07} Schneider, D.P. et al.\ 2007, \aj, 134, 102

\bibitem[Spergel et al.\ (2007)]{spe07} Spergel, D.N. et al.\ 2007, \apjs, 170, 377

\bibitem[Steffen et al.\ (2006)]{ste06} Steffen, A.T., Strateva, I., Brandt, W.N., Alexander,
	D.M., Koekemoer, A.M., Lehmer, B.D., Schneider, D.P., \& Vignali, C. 2006, 131, 2826

\bibitem[Tang et al.\ (2007)]{tan07} Tang, S.M., Shuang, N.Z., and \& Hopkins, P.F. 2007,
	\mnras, 377, 1113

\bibitem[Vanden Berk et al.\ (2001)]{van01} Vanden Berk, D.E. et al.\ 2001, \aj, 122, 549

\bibitem[Warren, Hewett \& Osmer (1994)]{war94} Warren, S.J., Hewett, P.C., \& Osmer,
	P.S. 1994, \apj, 421, 412

\bibitem[Webster et al.\ (1995)]{web95} Webster, R.L., Francis, P.J., Peterson, B.A., 
	Drinkwater, M.J., \& Masci, F.J. 1995, \nat, 375, 469

\bibitem[Wisotzki (1998)]{wis98} Wisotzki, L. 1998, {\it Astron. Nachr.}, 319, 257

\bibitem[Wisotzki (2000)]{wis00} Wisotzki, L. 2000, \aap, 353, 861 

\end{thebibliography}
\end{document}